\documentclass[aps,prc,superscriptaddress,showpacs,onecolumn]{revtex4}	
 \usepackage{amsmath}  
 \usepackage{makeidx}
 \usepackage{graphicx}
 \usepackage{amsfonts}
 \usepackage[ansinew]{inputenc}
 \usepackage{subfigure}
 \usepackage{color}
 \usepackage{mathrsfs}

              \makeindex

\newcommand{\lsim}{\stackrel{\scriptstyle <}{\phantom{}_{\sim}}}

\def\eqs{\makebox(0,0){$=$}}
\def\pls{\makebox(0,0){$+$}}

\def\ssp{\makebox(0,0)
    {\thinlines\put(-.1,0){\line(1,0){.2}}\put(0,-.1){\line(0,0){.2}}}}
\def\ssm{\makebox(0,0){\put(-.1,0){\thinlines\line(1,0){.2}}}}
\def\photon{\thinlines\multiput(0,0)(.2,0){3}{\line(1,0){0.1}}}

\def\oneloop{
     \put(1.5,0){\thicklines\oval(2.0,1.5)}
     \put(0,0){\photon}\put(0.3,0.3){\ssp}
     \put(2.5,0){\photon}\put(2.7,0.3){\ssm}}
\def\fullself{\begin{picture}(5,1)\put(0,0){\oneloop}
     \put(1.5,0){\makebox(0,0){$-\ii{\boldsymbol \Pi}$} }
     \end{picture}}

\def\oneloopvertex{
    \put(1.625,0){\thicklines\oval(2.0,1.5)}
    \put(0,0){\photon}\put(0.35,0.3){\ssp}
    \put(0.625,0){\circle*{.25}}\put(2.625,0){\circle*{.25}}
    \put(2.75,0){\photon}\put(2.9,0.3){\ssm}}

\def\fullbox{\makebox(0,0){\rule{2mm}{4mm}}}

\def\interaction{\makebox(0,0){\put(0,0){\interact}
    \put(0,.95){\ssp}\put(0,-.95){\ssm}
    \put(0,.125){\ssp}\put(0,-.125){\ssm}}}
\def\interact{\makebox(0,0){\put(0,.5){\fullbox}
    \thicklines\put(0,0){\oval(.75,.5)}
    \put(0,-.5){\fullbox}} }

\def\til2loop{\put(0,0){\oneloopvertex}\put(4.,0){\pls}
      \put(4.75,0){\oneloopvertex}\put(6.375,0){\interaction}
      \put(9,0){\pls}}

\newcommand{\di}{{\mathrm d}}
\newcommand{\ii}{{\mathrm i}}

\begin{document}
\title{Fermi liquid, clustering, and structure factor in dilute warm nuclear matter}
\author{G. R\"{o}pke}
\email{gerd.roepke@uni-rostock.de}
\affiliation {Institut f\"{u}r Physik, Universit\"{a}t Rostock,  Rostock, Germany}
\affiliation {National Research Nuclear University (MEPhI),  Moscow, Russia}
\author{D.N. Voskresensky}
\affiliation {National Research Nuclear University (MEPhI),  Moscow, Russia}
\affiliation {Joint Institute for Nuclear Research (JINR), Russia}
\author{I.A. Kryukov }
\affiliation {National Research Nuclear University (MEPhI),  Moscow, Russia}
\author{D. Blaschke }
\affiliation {National Research Nuclear University (MEPhI),  Moscow, Russia}
\affiliation {Joint Institute for Nuclear Research (JINR), Russia}
\affiliation {Institute of Theoretical Physics, University Wroclaw, Poland}
\date{\today}

\begin{abstract}
Properties of nuclear systems at subsaturation densities can be obtained from different approaches.
We demonstrate the use of the density autocorrelation function which is related to the isothermal compressibility
and, after integration, to the equation of state.
This way we connect the Landau Fermi liquid theory well elaborated in nuclear physics with the
 approaches to dilute nuclear matter describing cluster formation.
A quantum statistical approach is presented, based on the cluster decomposition of the polarization 
function.
The fundamental quantity to be calculated is the dynamic structure factor.
Comparing with the Landau Fermi liquid theory which is reproduced in lowest approximation,
the account of bound state formation and continuum correlations gives the correct low-density result as
described by the second virial coefficient and by the mass action law (nuclear statistical equilibrium).
Going to higher densities, the inclusion of medium effects is more involved compared with other quantum statistical
approaches, but the relation to the Landau Fermi liquid theory gives a promising approach to describe not only thermodynamic
but also collective excitations and  non-equilibrium properties of nuclear systems in a wide region of the phase diagram.
 \end{abstract}


\maketitle

 \section{Introduction}
\label{Sec:Introduction}

The nuclear matter equation of state (EoS) was investigated recently with respect to various applications such as astrophysics of compact stars, the structure of nuclei, and heavy ion collisions (HIC).
The derivation of the EoS from a microscopic description has several difficulties:
One has to deal with a strongly interacting fermion system with arbitrary degeneracy.
Correlation effects, such as bound state formation and quantum condensates, have to be treated.
Effective nucleon-nucleon interactions are used, which are reconstructed from measured properties
and are possibly dependent on density and energy.
Different approximations are known which  describe nuclear matter in limiting cases,
such as at low densities or at near-saturation density,
but a theory applicable in a wide region of temperature and density is needed.
Often, the processes under consideration are in non-equilibrium and are spatially inhomogeneous
so that the assumption of local thermodynamic equilibrium becomes questionable.

We investigate nuclear matter in thermodynamic equilibrium, confined in the volume $V$ at temperature $T$.
We are interested in the subsaturation region where the baryon density $n=n_n+n_p\leq n_{\rm sat}$
with the saturation density $n_{\rm sat} \approx 0.15$ fm$^{-3}$,
the temperature $T \lsim 20$ MeV,
and the proton fraction $Y_p=n_p/n$ between 0 and 1.
As long as weak processes leading to $\beta$-equilibrium are suppressed, the number of neutrons and protons
$N_\tau = n_\tau V$ are independent variables,  $\tau =n,p$.
 This region of warm dense matter is of interest  for nuclear structure calculations
and HIC
explored in laboratory experiments \cite{Natowitz}, and also for astrophysical applications.
For instance,  core-collapse supernovae at the post-bounce stage are evolving within this region of the phase diagram \cite{Tobias},
and different processes such as  neutrino emission and absorption, which strongly depend on the composition of warm dense matter,
influence the mechanism of supernova explosions.

Different approaches have been worked out to describe warm 
nuclear matter.
At low densities, the formation of bound states is of relevance. The simple mass action law \cite{RMS,NSE}
gives already the nuclear statistical equilibrium (NSE). The exact behavior of the EoS
 in the low-density limit is obtained
from the second virial coefficient \cite{BU,RMS,SRS,HS}. With increasing density, medium effects must be considered.
Other approaches have been worked out to treat dense nuclear systems.
For instance, the relativistic mean-field (RMF) approach \cite{Walecka} has been widely recognized to be powerful
in describing nuclear systems.
Properties near the saturation density are used to fix the parameters of the RMF approach,
see, e.g., Ref.~\cite{Typel1999}.

An alternative approach to describe the properties of nuclear matter near the saturation density
is based on the theory of normal Fermi liquids built up by Landau \cite{L56} which is designed for strongly degenerate systems.
The application of the Fermi-liquid approach to nuclear systems was developed by Migdal \cite{Mig}, see also \cite{Mig1}.
The approach describes low-lying excitations by several phenomenological Landau-Migdal  parameters.
Pomeranchuk \cite{Pom58} has shown that Fermi liquids are stable only if some inequalities on the values of the Landau-Migdal parameters are fulfilled.
In a recent work \cite{KV16}, low-lying scalar excitation modes
in cold normal Fermi liquids have been investigated
for various values
of the scalar Landau-Migdal parameter $f_0$ in the particle-hole channel.
The stability of nuclear matter was then discussed.
After performing the bosonization of the local interaction, the possibility of Bose condensation of scalar quanta  has been suggested that may result in the appearance of a novel metastable state in dilute nuclear matter.

The RMF approach as well as the Landau Fermi-liquid approach are based on a single-nucleon quasiparticle concept. In both cases it is not simple to introduce the formation of clusters, in particular of light elements with mass number $A \le 4$.
A generalization of the RMF approach has been proposed \cite{Typel,ferreira12,PCP15},
where light elements $^2$H ($d$), $^3$H ($t$), $^3$He ($h$), $^4$He ($\alpha$) are considered as new degrees of freedom in the effective Lagrangian.
The coupling constants of the light clusters to the meson fields are adapted from other theories,
in particular the quantum statistical (QS) approach.
Within the generalized RMF approach, the second virial coefficient is reproduced \cite{VT} fitting special terms in the EoS.

A similar problem arises also in the Landau Fermi-liquid approach. In Fig. 5 of Ref. \cite{KV16}, the energy as function of the baryon density is shown, taking into account the possibility of the Bose condensation of the scalar quanta.
Compared with the result of the original Fermi liquid theory, at low densities the energy might be shifted downwards due to Bose condensation. However, clustering is not included and the low-density virial expansion for finite temperature  is not correctly reproduced. Until now, no systematic approach is known how to incorporate bound state formation in the Landau Fermi-liquid approach.

In contrast to these single-nucleon mean-field theories, a QS approach \cite{RMS} describes the formation of clusters
and their dissolution at increasing densities (Mott effect) in a systematic way.
Light elements are considered as quasiparticles in the corresponding few-nucleon spectral function.
Medium-dependent quasiparticle energies are given in Ref. \cite{R}.
The corresponding EoS \cite{Typel} interpolates between both limiting cases, the mean-field approach near the saturation density, and the low-density region where clusters are significant.

In the present work we will treat the problem how the Landau Fermi-liquid approach can be improved to include the formation of clusters.  Although
the Fermi-liquid approach can be applied to non-equilibrium and inhomogeneous systems and to the systems with Cooper pairing, cf. \cite{Voskresensky:1993ud,Kolomeitsev:2010pm},
within this work we focus on equilibrium properties of the normal homogeneous nuclear matter.
We show in our dynamic structure factor approach that the low-density limit is correctly described. Formation of light clusters, in particular the second virial coefficient, are reproduced.
In detail, we\\
(i) investigate the polarization function (Fourier transform of the van Hove time-dependent pair correlation function) in relation to the EoS and establish the connection of the QS approach with the Fermi-liquid approach,\\
(ii) consider the dynamic structure factor as the central quantity which contains not only thermodynamic information, but also transport and kinetic properties,\\
(iii) introduce the formation of clusters and the inclusion of continuum correlations, \\
(iv) discuss the stability with respect to phase transitions.\\
Bosonization and Bose condensation together with metastability \cite{KV16} will be considered in a subsequent work.

We compare in this work two different approaches to the thermodynamical properties of nuclear matter: 
Firstly, we consider the density equation of state based on the single-particle Green function, 
and secondly the isothermal compressibility related to the dynamic structure factor, related to the density-density correlation function. 
We investigate the applicability of the quasiparticle approximation (QPA) and the Fermi-liquid approach, 
and compare results for the relativistic mean-field approximation with those of  the Fermi-liquid approach. 
Then, cluster formation is treated, and the Beth-Uhlenbeck formula is obtained in dilute warm nuclear matter 
as well as the isothermal compressibility is considered. 
Accurate results for the incompressibility are given valid in the low-density limit.

The paper is organized as follows.
In Sec. \ref{Isothermal} we elaborate the fluctuation-dissipation theorem.
 Simplifying we consider cases of  pure neutron matter and isospin-symmetric matter.
We relate the dynamic
 and static structure factors, expressed in terms of
diagrams in the Matsubara or the Schwinger-Kadanoff-Baym-Keldysh techniques,
to the isothermal compressibility.
In Sec. \ref{Examples} and appendix \ref{sec:spectral} we consider special approximations, in particular the limiting cases of the perturbative loop diagram,
the one -- loop contribution with full and quasiparticle Green functions, the RMF approximation,
and the RPA resummation.
In Sec. \ref{sec:Fermi} we derive the static structure factor within the Fermi-liquid approach.
In Sec. \ref{RMF} we describe density-density correlations within the RMF approximation and recover the Landau-Migdal parameter in the scalar channel.
Generalizations to systems with arbitrary isospin composition are performed in Sec. \ref{arbitrary}.
Then in Sec. \ref{two} we study two-particle correlations.
We present the Beth-Uhlenbeck approach for the second virial coefficient and discuss the cluster decomposition of the polarization function.
In Sec. \ref{sec:NSE} we show how the nuclear statistical equilibrium model appears in the low-density limit.
We discuss the corrections at increasing density and give concluding remarks in Sec. \ref{Conclusions}.
In this work we use units  $\hbar =c=k_{\rm B}=1$.

\section{Isothermal compressibility and the dynamic structure factor}
\label{Isothermal}

\subsection{The fluctuation-dissipation theorem}
\label{FDT}

Starting from a given nucleon-nucleon interaction potential, quantum statistics gives the possibility to derive thermodynamic potentials and related thermodynamic properties.
Using the technique of thermodynamic  Green functions \cite{AGD},
different approaches are known to calculate an EoS.
For instance, one can evaluate expressions  for the pressure $p(T,\mu_n,\mu_p)$,
or for the  neutron and proton densities $n_\tau (T,\mu_n,\mu_p)$ for nuclear matter \cite{R}.
Here, $\mu_n, \mu_p$ denote the chemical potentials of neutrons and protons, respectively; $\tau =\{n,p\}$.
For instance, one can start from the expression for the
density of the fermion species (density EoS)
\begin{equation}
\label{neos}
n_\tau(T,\mu_n,\mu_p)= \frac{1}{V}\sum_{{\bf p},\sigma}\int_{-\infty}^\infty \frac{d \omega}{2 \pi}f(\omega-\mu_\tau)
A_\tau(\omega,{\bf p}; \mu_n,\mu_p)
\end{equation}
with the Fermi function
\begin{equation}
\label{Fermi}
f(z) =\frac{1}{e^{z/T}+1}
\end{equation}
and the single-particle spectral function $A_\tau(\omega,{\bf p}; \mu_n,\mu_p)$, see  \cite{AGD}.  In infinite matter we replace the sum over the single-particle states $p=\{ {\bf p},\sigma\}$ by $g_\tau V \int d^3p/(2 \pi)^3$, where $g_\tau$ is the spin degeneracy factor, $\bf p$ is the wave number vector.
Within the Matsubara Green function method, the spectral function is related to the self-energy, and systematic
approaches are available to calculate this EoS in an appropriate approximation.
This way, thermodynamic potentials have been derived for nuclear systems and  further thermodynamic properties including  composition, phase transitions, etc., have been considered.
Note that in the case of phase transition not only the baryon density $n$, but also the proton fraction $Y_p$ may be different in the different phases.  The total baryon number is conserved, in our case $N=N_p+N_n$.

A related approach is based on the real time non-equilibrium Green function technique
\cite{KB,Keldysh}.
The Wigner transformed Green functions and corresponding self-energies
allow to describe slightly non-uniform systems involved in slow  collective motions.
The technique is also convenient to describe systems in thermal equilibrium. Here,   the Wick rotation is not required. All the non-equilibrium Green functions and self-energies are expressed via the retarded quantities.
General issues of the quantum kinetic approach were reviewed in \cite{Kolomeitsev:2013du}.
In  Ref. \cite{Voskresensky:1987hm} the QPA in thermal equilibrium was applied within the non-equilibrium Green function technique. Within the QPA for the
nucleon Green functions the formalism has been applied to special problems in nuclear systems, cf. \cite{Voskresensky:2001fd}. The general formalism  developed in \cite{Knoll:1995nz}
allows to account for the finite damping width of the
source particles due to their finite mean free path in matter.

To simplify the consideration we first consider the case of one baryon species that holds for the pure neutron matter (degeneracy factor $g=2$) and isospin-symmetric matter (degeneracy factor $g=4$ including isospin). Generalizations for systems with arbitrary isospin composition will be presented below.
In the present work,
the inclusion of few-nucleon correlations and the possible formation of bound states
is of interest. To evaluate thermodynamic properties of nuclear systems,
we investigate the isothermal compressibility $\kappa_{\rm iso}$,
\begin{equation}
\label{kappa}
\kappa_{\rm iso}(T,\mu)=-\frac{1}{V}\frac{\partial V}{\partial p}{\Big |}_T=
\frac{1}{n^2} \frac{\partial n}{\partial \mu}{\Big |}_T\,,
\end{equation}
where $\mu$ is the baryon chemical potential.
If this quantity is known, we can derive the EoS: we integrate
$n(T,\mu)$
at fixed
$T$ to obtain the pressure
\begin{equation}
\label{pressure}
p(T,\mu)=\int_{-\infty}^\mu  n(T,\mu')d \mu'
\end{equation}
as thermodynamic potential.
The incompressibility is defined as
\begin{equation}
\label{incompr}
K(T, n)=\frac{1}{n\, \kappa_{\rm iso}}=n \frac{\partial \mu}{\partial n}\Big|_T.
\end{equation}

The isothermal compressibility is related to another fundamental quantity of the many-body
system, the  dynamic structure factor.
\begin{equation}
S({\bf q},\omega)=\frac{1}{2 \pi V} \int_{-\infty}^\infty dt \langle \rho_{\bf{q}}^+(t) \rho_{\bf q}(0)\rangle e^{\ii \omega t}\,.
\end{equation}
Here, the wave number dependent density fluctuation
\begin{equation}
\label{densityrhoq}
\rho_{\bf q}=\int d^3r\, e^{\ii {\bf q} \cdot {\bf r}} \sum_\nu\psi_\nu^\dagger({\bf r}) \psi_\nu({\bf r})
 \end{equation}
 is the Fourier transform of the particle density. The intrinsic quantum numbers are denoted by
$\nu = \{\sigma, \tau\}$, and
$\psi_\nu^\dagger({\bf r}),\psi_\nu({\bf r})$ are the corresponding creation and annihilation operators.
The time dependence is according to the Heisenberg picture. The average $\langle \dots
\rangle={\rm Tr}\{ \rho_{\rm eq} \dots \}$ is performed with the equilibrium statistical operator, in our case the grand canonical statistical operator
\begin{equation}
\rho_{\rm eq}(T, \mu_\nu)
=\frac{e^{-(H-\sum \mu_\nu N_\nu)/T}}{{\rm Tr}\,e^{-(H-\sum \mu_\nu N_\nu)/T}},
 \end{equation}
with the particle number operators $N_\nu=\int  d^3r \,
\psi_\nu^\dagger({\bf r}) \psi_\nu({\bf r})$. 

From the dynamic structure factor, the static structure factor,
\begin{equation}
 S({\bf q})=\int \frac{d \omega}{2 \pi}\, S({\bf q},\omega),
\end{equation}
 is derived,
describing equal-time fluctuations of the baryon density
$\hat n({\bf r})=\sum_\nu\psi_\nu^\dagger({\bf r}) \psi_\nu({\bf r})$.
The generalization to a multicomponent system where partial structure factors can be introduced, will be discussed below, see Eq. (\ref{partSF}).
 In equilibrium systems where $n=\langle \hat n({\bf r})
\rangle$,  the static structure factor can be rewritten in terms of the density derivative of the baryon chemical potential at fixed temperature \cite{LL1980}, Sect. XII,
\begin{eqnarray}\label{EqStstr0}
S({\bf q}\to 0)=(\langle \hat n^2({\bf r})
\rangle-n^2)/n=T\left\{\left(\frac{\partial \mu}{\partial n}\right){\Big |}_T\right\}^{-1}\,.
\end{eqnarray}

There are other quantities which are related to
$S({\bf q},\omega)$
so that the investigation of this quantity is of fundamental interest.
According to the fluctuation-dissipation theorem, the dynamic structure factor
is related to the response function $\chi({\bf q},\omega)$ defined by the density fluctuation induced by an external potential
$U({\bf q},\omega)$ as $\langle \rho_{\bf q}(\omega) \rangle=\chi({\bf q},\omega) U({\bf q},\omega)$.
The response function is related to the thermodynamic density-density Green function
\begin{equation}
L(1,2;1^+,2^+) = \frac{V}{\ii^2}
\langle {\rm T}\{\psi^\dagger(1^+)\psi(1) \psi^\dagger(2^+) \psi(2) \}\rangle-
\frac{V}{\ii^2} \langle \psi^+(1^+)\psi(1) \rangle \langle \psi^+(2^+) \psi(2) \rangle\,,
\end{equation}
 with $1 = \{{\bf r}_1,\sigma_1,\tau_1\}= \{{\bf r}_1,\nu_1\}$. As well-known from the technique
of thermodynamic Green functions \cite{KB,AGD}
we introduce the Heisenberg-like dependence on the parameter $x$ according to
\begin{equation}
A(x)= e^{x (H-\sum \mu_\nu N_\nu)}Ae^{-x (H-\sum \mu_\nu N_\nu))};
\end{equation}
the ${\rm T}\{\dots \}$-product denotes the ordering of operators
with growing parameter values $x$
from right to left. After Fourier transformation with respect to spatial distances and the
parameter $x$,
\begin{equation}
\label{Lqo}
 L({\bf q},\ii z_\lambda)=\int d^3 r\int_0^{1/T} d x
e^{\ii {\bf q} \cdot ({\bf r}_2-{\bf r}_1)}
e^{\ii z_\lambda (x_2-x_1)} L(1,2;1^+,2^+)
\end{equation}
is a function defined at the bosonic Matsubara frequencies $z_\lambda=\pi \lambda T$,
where $\lambda = 0,\,\, \pm 2, \dots$ are the even numbers.
Analytical continuation from $\ii z_\lambda$ into the whole complex $z$-plane
gives the corresponding damping width (spectral function of the van Hove-function $L$)
\begin{equation}
\label{Spectral}
 \Gamma_{L} ({\bf q},\omega)=\ii[L({\bf q},\omega+\ii 0)-L({\bf q},\omega-\ii 0)]
=2 \,{\rm Im}\,L({\bf q},\omega-\ii 0)\,.
\end{equation}
According to the fluctuation-dissipation theorem,
the response function $\chi({\bf q},\omega)$ which describes also transport and absorption
 is related to density fluctuations according to
\begin{equation}
{\rm Im}\chi({\bf q},\omega)=-{\rm Im}L({\bf q},\omega-\ii 0).
\end{equation}
As immediately seen from the spectral representation, the density-density correlation function is  obtained from the width-function after multiplication with the Bose factor,
here
\begin{equation}\label{SL}
S({\bf q},\omega)=\frac{1}{\pi} \frac{1}{e^{ \omega/T}-1}{\rm Im}L({\bf q},\omega-\ii 0).
\end{equation}
For the  isothermal compressibility (\ref{kappa}) we find
\begin{eqnarray}\label{isot}
&&\kappa_{\rm iso}(T,\mu)
=\frac{1}{
 n^2T} \lim_{{\bf q} \to 0} \int_{-\infty}^\infty \frac{d \omega}{\pi}
\frac{1}{e^{ \omega/T}-1}{\rm Im}L({\bf q},\omega-\ii 0).
\end{eqnarray}
 For $L({\bf q},\ii z_\lambda)$ a systematic perturbation expansion with the Matsubara technique is available which can be represented by Feynman diagrams.
Below we show how bound state formation is treated this way.

\subsection{Real-time Green function technique}
\label{sec:nonequGF}

Another quantum statistical approach to calculate the isothermal compressibility (\ref{kappa}) starts from the real time non-equilibrium Green
function technique.
The four-dimensional nucleon current-current
auto-correlation function \cite{Voskresensky:1987hm,Knoll:1995nz} is given by the $-+$
component of their Wigner transformed self-energy
\begin{eqnarray} \label{Pi-+xq}\unitlength6mm
   -\ii\Pi^{\mu\nu}_{-+}(q;X) = \begin{picture}(4.3,.7) \put(0.2,0.2){\fullself}
   \put(3.8,0.2){\eqs} \end{picture} \int \di^4 \xi e^{\ii
   q\xi}\langle j^{\mu\dagger}(X-\xi/2) j^\nu(X+\xi/2)\rangle,
\end{eqnarray}
with $q^\alpha =(\omega, {\bf q})$, $X^\alpha =(x^\alpha +y^\alpha)/2$, $\alpha =0,1,2,3$. The dashed lines relate to a vector boson interacting with nucleons by a boson-two-fermion interactions, like in QED, while the
$-\ii\Pi $-loop symbolically denotes the exact inclusion of all strong
interactions among the source particles. The bracket
$\langle\dots\rangle$ denotes a quantum ensemble average over the
source with quantum states and operators in the interaction picture.
The $\Pi_{-+}(q; X)$ and $ \Pi_{+-}(q; X)$ self energies have meaning of
the gain and loss terms in the Kadanoff-Baym quantum kinetic equation.
We use the convenient up and down $\{-,+\}$ contra-variant and co-variant notations of
\cite{Ivanov:1999tj}. They are introduced in Appendix \ref{A1} as well as  relations between equilibrium two-point functions.

In the non-relativistic case considered here, the nucleon density-density correlator is determined as $-\ii\Pi^{00}_{-+}$, calculated with $j_\mu =(\hat n, {\bf 0})$, $\hat n$ is the density of the fermion species under consideration,  the bare vertices are taken as  $V_0^{\pm} =(1, {\bf 0})$, $V_{0,\pm} =(\pm 1, {\bf 0})$. The Wigner
  transform of $-\ii\Pi^{00}_{-+}(x,y)$ has the meaning of the dynamic structure factor, \begin{eqnarray}
-\ii\Pi^{00}_{-+}(q;X)=S(q;X)\,.
\end{eqnarray}
The static structure factor is then as follows
\begin{eqnarray} \label{stSpectr}
S({\bf q}\to 0; X)=(\langle \hat n^2 \rangle-\langle \hat n \rangle^2)/\langle \hat n \rangle=\int_{-\infty}^{\infty} \frac{\di q_0}{2\pi}[-\ii\Pi^{00}_{-+}(q_0,{\bf q}\to 0; X)].
\end{eqnarray}
It has the meaning of the normalized variance of the fermion density in the medium.

In thermodynamic equilibrium, the dependence on $X$ disappears because of homogeneity
in space and time. Applying the first relation (\ref{eqrel}), cf. (\ref{EqStstr0}), we get the exact relation
\begin{eqnarray} \label{EqstSpectr}
S({\bf{q}}\to 0)=\lim_{{\bf{q}}\to 0}\int_{-\infty}^{\infty} \frac{\di q_0}{2\pi}\frac{\Gamma_{\rm B}(q_0,{\bf{q}})}{e^{q_0/T}-1}\,, \quad\Gamma_{\rm B}(q_0,{\bf{q}})=-2\mbox{Im}\Pi^{ R}(q_0,{\bf{q}})\,,
\end{eqnarray}
where $\Pi^R$ is the boson retarded self-energy, and $q_0=\omega$ as used above.

The boson width, $\Gamma_{\rm B}(q_0,{\bf{q}})$, in (\ref{EqstSpectr}) is non-zero only in the space-like region $q_0<|{\bf{q}}|$ and thus
\begin{eqnarray} \label{0EqstSpectr}
S({\bf{q}}\to 0)=T\lim_{{\bf{q}}\to 0}\int_{-\infty}^{\infty} \frac{\di q_0}{2\pi}\frac{\Gamma_{\rm B}(q_0,{\bf{q}})}{q_0}=T\mbox{Re}\Pi^R (0,{\bf{q}}\to 0) \,,
\end{eqnarray}
where we used the corresponding Kramers-Kronig relation.

Also, as follows from the definition of $G^{-+}$ fermion Green function, $\ii G^{-+}=-\langle \psi_2^\dagger \psi_1\rangle$ and  (\ref{neos}), (\ref{eqrel})  the fermion density is given by (see Eq. (\ref{neos}) and below)
\begin{eqnarray}\label{EQdistr}
 n_\tau=g_\tau\int\frac{\di^4 p}{(2\pi)^4} \frac{A_\tau(p)}{e^{(p_0-\mu)/T}+1}\,,\quad A_\tau(p)=-2\mbox{Im}G^R_\tau(p)\,,
\end{eqnarray}
where $A_\tau(p)$ is the single-particle spectral function
which obeys the sum-rule 
\begin{eqnarray}
\int_{-\infty}^{\infty} A_\tau( p) \frac{dp_0}{2\pi}=1\,,\quad \tau =n,p\,.
\end{eqnarray}
For pure neutron matter ($g=2$), the chemical potential $\mu=\mu_n$ is given by the neutron chemical potential. For isospin-symmetric matter, the total density is $n=n_n+n_p=2 n_n$ since the chemical potentials of neutrons and protons coincide, $\mu=\mu_n=\mu_p$, in total $g=4$. To get
coincidence with Eq. (\ref{neos}) we performed the variable  shift $p_0\to p_0 -\mu$, so that the Green functions depend now on $p_0$ and occupations, on $p_0 -\mu$.
In the presence of the vector field $V=(V_0, {\bf V})$, like for the nucleon interacting with the $\omega$ meson, we should still perform the shift $\mu\to\mu-V_0$.

Concluding, first we  used the Matsubara technique
and then we applied the non-equilibrium diagram technique. We see that  the quantity  $L({\bf{q}},\omega-\ii 0)$ appeared in (\ref{Spectral}), (\ref{SL}) has the meaning of the advanced self-energy $\Pi^A$,  $L({\bf q},\omega+i 0)=\Pi^R(q_0,{\bf{q}})$, $\Gamma_L =\Gamma_{\rm B}=-2\mbox{Im}\Pi^{R}(q_0,{\bf{q}})$
 and
\begin{equation}
-\ii \Pi^{00}_{-+}(\omega,{\bf q})
= \frac{\Gamma_L({\bf q},\omega)}{e^{\omega/T}-1} = \frac{2{\rm Im}\,L({\bf q},\omega-i 0)}{e^{ \omega/T}-1}  \,.
 \end{equation}
The real parts follow from the Kramers-Kronig relation.
In the long-wavelenth limit, we recover Eq. (\ref{isot}).

\section{Various  approximations}
\label{Examples}

\subsection{Perturbation expansion: lowest order}

\begin{figure}[!th]
\includegraphics[width=6cm]{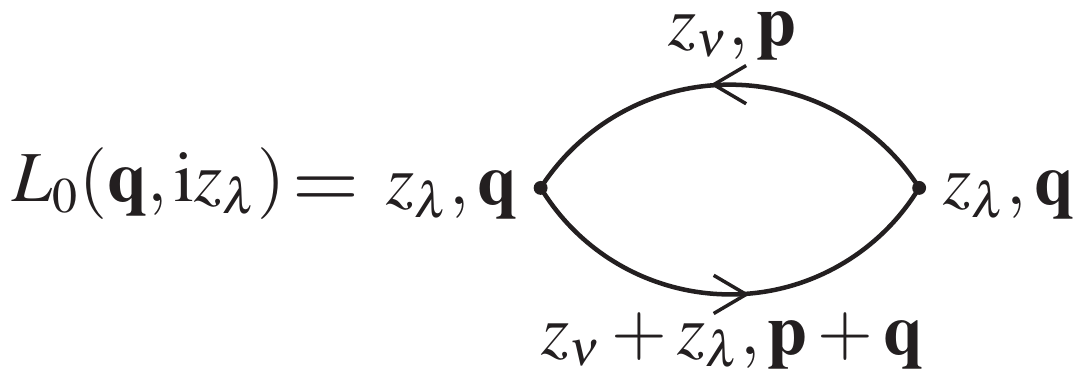}
\caption{The one-loop perturbative contribution to the  polarization function in Matsubara technique, $L_{0}({\bf q},\ii z_\lambda)$ (zeroth order of interaction). }
\label{fig:1}
\end{figure}

To become more familiar with the formalism we briefly discuss the trivial case of the
lowest order  perturbation theory where all interactions are neglected,
and the ideal non-relativistic Fermi gas results.
In the simplest approximation (zeroth order of the interaction) we have the single-particle polarization loop result (see Fig. \ref{fig:1})
\begin{equation}\label{Lo}
L_0({\bf q},z)=
g \int \frac{d^3 p}{(2\pi)^3}\frac{f(\epsilon^0_p)-f(\epsilon^0_{{\bf p}+{\bf q}})
}{z+\epsilon^0_{{ p}}-\epsilon^0_{{\bf p}+{\bf q}}}
\end{equation}
with $\epsilon^0_{{\bf p}}=\epsilon^0_p=E^0_p-\mu$, with  $E^0_p=p^2/(2m)$ (the rest energy term is shifted to $\mu$).
For simplicity the index $\nu$ is dropped, the generalization from single-component to multi-component matter is trivial.

Using (\ref{SL}), (\ref{Lo}) we get the result for the lowest order of perturbation theory
\begin{equation}
\label{S0q}
S_0({\bf q},\omega)= \frac{g}{e^{ \omega/T}-1}
\int \frac{d^3 p}{(2\pi)^3}\left[f(\epsilon^0_p)-f(\epsilon^0_{{\bf p}+{\bf q}})\right] \delta[\omega+\epsilon^0_{{ p}}-\epsilon^0_{{\bf p}+{\bf q}}]\,.
\end{equation}
With
\begin{equation}
\label{f1-f}
 f(\epsilon^0_p)-f(\epsilon^0_{{\bf p}+{\bf q}})=
f(\epsilon^0_{{\bf p}+{\bf q}})\left[1-f(\epsilon^0_p)\right]
\left(e^{(\epsilon^0_{{\bf p}+{\bf q}}-\epsilon^0_{{p}})/T}-1 \right)
\end{equation}
we obtain for the resulting isothermal compressibility (\ref{isot}) after integration over $\omega$
\begin{eqnarray}
\label{eq:22}
&&
\kappa^{(0)}_{\rm iso}(T,\mu)
=\frac{g}{n^2T}
 \int \frac{d^3 p}{(2\pi)^3}f(\epsilon^0_p) [1-f(\epsilon^0_p)].
\end{eqnarray}
The integral
is performed using integration by parts
\begin{eqnarray}
\int \frac{d^3 p}{(2\pi)^3}f(\epsilon^0_p)  [1-f(\epsilon^0_p) ]=
-\frac{4 \pi}{(2 \pi)^3}\int_0^\infty p^2dp\frac{m T}{  p} \frac{df(\epsilon^0_p) }{dp}
=\frac{ m T}{2\pi^2  }\int_0^\infty dp\,f(\epsilon^0_p).
\end{eqnarray}

The same resulting expression
\begin{equation}
\kappa^{(0)}_{\rm iso}(T,\mu)=
g \frac{m}{2\pi^2  n^2}\int_0^\infty dp\,f(\epsilon^0_p)
\end{equation}
is obtained from the EoS of the ideal quantum gases. Neglecting all interactions we have
\begin{equation}
n^{(0)}(T, \mu)=\frac{1}{V} \sum_{p}\frac{1}{e^{ \epsilon^0_p/T}+1}=\frac{1}{V} \sum_{p} f(\epsilon^0_p)
=\frac{N}{V}\,,
\end{equation}
and $\partial n^{(0)}/\partial \mu= 1/(V T)\sum_p f(\epsilon^0_p)  [1-f(\epsilon^0_p) ]$ in agreement with Eq. (\ref{eq:22}) with $\sum_p \to g V\int d^3p/(2 \pi)^3$. For symmetric matter, in addition to spin also isospin summation bas to be performed.  The introduction of quasiparticles within the Matsubara Green function approach is discussed below in Sec. \ref{sec:mean}.

The lowest order perturbation theory where all interactions are neglected, i.e. the ideal Fermi gas, is often used as a system of reference.
To investigate the influence of the interaction on the incompressibility $K(T,n)$ (\ref{incompr}) in  isospin-symmetric matter,
we introduce the excess quantity $\varphi_0(T,n)$ according to
\begin{equation}
\label{fi0}
 K(T,n)=K^{(0)}(T,n)[1+\varphi_0(T,n)]
\end{equation}
where $K^{(0)}(T,n)=1/[n\kappa_{\rm iso}^{(0)}(T,n)]$ according to Eq. (\ref{eq:22}). Results for $\varphi_0(T,n)$ are presented in Sec. \ref{comp}.

\subsection{The single-particle mean-field approximation}
\label{sec:mean}

To give a simple example for the quasiparticle approach,
we discuss the lowest order with respect to the interaction where the mean-field (Hartree-Fock, HF) approximation is obtained.  We start from the nonrelativistic Hamiltonian
\begin{equation}
 H=\sum_1 \epsilon^0_1 \hat{a}^\dagger_1\hat{a}_1+H_{\rm int}, \qquad H_{\rm int}=\frac{1}{2} \sum_{12,1'2'} V(1,2;1',2') \hat{a}^\dagger_{1'} \hat{a}^\dagger_{2'}\hat{a}_2\hat{a}_1.
\end{equation}
Here $\hat{a}^\dagger_1, \hat{a}_1$ are creation and annihilation operators,
respectively, for the single-nucleon state $1=\{{\bf p}_1,\nu_1\}$ denoting wave number, spin, and isospin.
As above we consider firstly effective one-component systems (neutron matter, symmetric matter) where the summation over spin and isospin is replaced by the degeneracy factor $g =2$ or 4, respectively, so that only the momentum $\bf p$ remains to characterize the single-nucleon state. With the antisymmetrized interaction
we have for wave number ${\bf p}$ the quasiparticle energy in HF approximation ($1 \to {\bf p}, 2 \to {\bf k}$, spin and isospin are not given explicitly, but the Hartree term leads to the factor $g$)
\begin{eqnarray}
\label{EHF}
&&\epsilon^{\rm HF}_p = \epsilon^0_p+\int \frac{d^3k}{(2 \pi)^3}  V({\bf p},{\bf k};{\bf p},{\bf k})_{\rm ex} f(\epsilon^{\rm HF}_k)
=\epsilon^0_p+\Delta_p^{\rm HF},  \\ && V({\bf p},{\bf k};{\bf p},{\bf k})_{\rm ex}=\sum_{\nu_k}^g \left[
V({\bf p},{\bf k};{\bf p},{\bf k})-V({\bf p},{\bf k};{\bf k},{\bf p})\delta_{\nu_k,\nu_p}\right]
\end{eqnarray}

As it is well-known from mean-field approximations, the self-energy $\Delta_p^{\rm HF}$ has no dependence on frequency $\omega$ so that
it is purely real. The spectral function follows as
$A^{\rm HF}(\omega , {\bf{p}})=2\pi \delta (\omega-\mu - \epsilon_{p}^0 -\Delta_p^{\rm HF})$.

The density EoS (\ref{neos})  relating $T, \mu$ to $n$,
reads in the HF QPA
\begin{equation}
\label{nHF}
n^{\rm HF}(T, \mu)=g \int \frac{d^3p}{(2 \pi)^3} \frac{1}{e^{\epsilon^{\rm HF}_p/T}+1}
=g \int \frac{d^3p}{(2 \pi)^3}f(\epsilon^{\rm HF}_p).
\end{equation}
We invert this relation so that the relation $\mu = \mu^{\rm HF}(T,n)$ is obtained.
>From this, the incompressibility (\ref{incompr})
\begin{equation}
\label{KTnB}
K(T, n)=n \frac{\partial  \mu^{\rm HF}(T,n)}{\partial n}\Big|_T
\end{equation}
follows after performing $\partial /\partial n$ in Eq. (\ref{nHF}),
\begin{equation}
1=\frac{g}{2\pi^2T}\int_0^\infty p^2dp f(\epsilon^{\rm HF}_p)[1-f(\epsilon^{\rm HF}_p)] \left[1-\frac{\partial \Delta_p^{\rm HF}}{\partial \mu}\right] \frac{\partial \mu^{\rm HF}}{\partial n}\Big|_T.
\end{equation}
We obtain
\begin{equation}
\label{muHFn}
\frac{\partial \mu^{\rm HF}}{\partial n}\Big|_T=\left\{\frac{g}{2\pi^2 T}\int_0^\infty p^2dp 
f(\epsilon^{\rm HF}_p)[1-f(\epsilon^{\rm HF}_p)]
 \left[1- \int \frac{d^3k }{(2 \pi)^3T}V({\bf p},{\bf k};{\bf p},{\bf k})_{\rm ex}
 f(\epsilon^{\rm HF}_k)[1-f(\epsilon^{\rm HF}_k)] \right]\right\}^{-1}.
\end{equation}

At low temperatures we replace ($k_{\rm F}=(6 \pi^2 n/g)^{1/3}$)
\begin{equation}
\label{Fermiint}
 f(\epsilon^{\rm HF}_k)[1-f(\epsilon^{\rm HF}_k)]
  =-\frac{\partial  f(\epsilon^{\rm HF}_k)}{\partial k} \frac{m^* T}{k}
\approx \delta(k-k_{\rm F})\frac{m^* T}{  k},
\end{equation}
where $m^*=(d^2 \epsilon_p/dp^2)$ at $p=k_{\rm F}$ is the non-relativistic effective mass of the quasiparticle  (the so called Landau effective mass).
We introduce the angular averaged  interaction
\begin{equation}
\tilde V(k_F)=(1/2)\int_{-1}^1 dz
V(p_{\rm F}{\bf e}_p,k_{\rm F}{\bf e}_k;p_{\rm F}{\bf e}_p,k_{\rm F}{\bf e}_k)_{\rm ex}
\end{equation}
with the unit vector scalar product ${\bf e}_p \cdot {\bf e}_k =z$ (isotropic interaction).
In first order we find
\begin{equation}
\label{KHF}
K^{\rm HF}=n \frac{\partial \mu^{\rm HF}}{\partial n}=\frac{\hbar^2k_{\rm F}^2}{3m}
\left[1+\frac{m^*k_{\rm F} }{2 \pi^2
}\tilde V(k_{\rm F})\right].
\end{equation}

Now we show the alternative way to obtain the EoS in the HF QPA
starting from the dynamic structure factor.
The thermodynamic density-density Green function $L({\bf q},\ii z_\lambda)$ (\ref{Lqo})
contains in addition to the zeroth order with respect to the interaction (\ref{Lo}) the following contributions
$L_1({\bf q}, \ii z_\lambda)$ which are of first order with respect to the interaction:
\begin{figure}[!th]
\includegraphics[width=11cm]{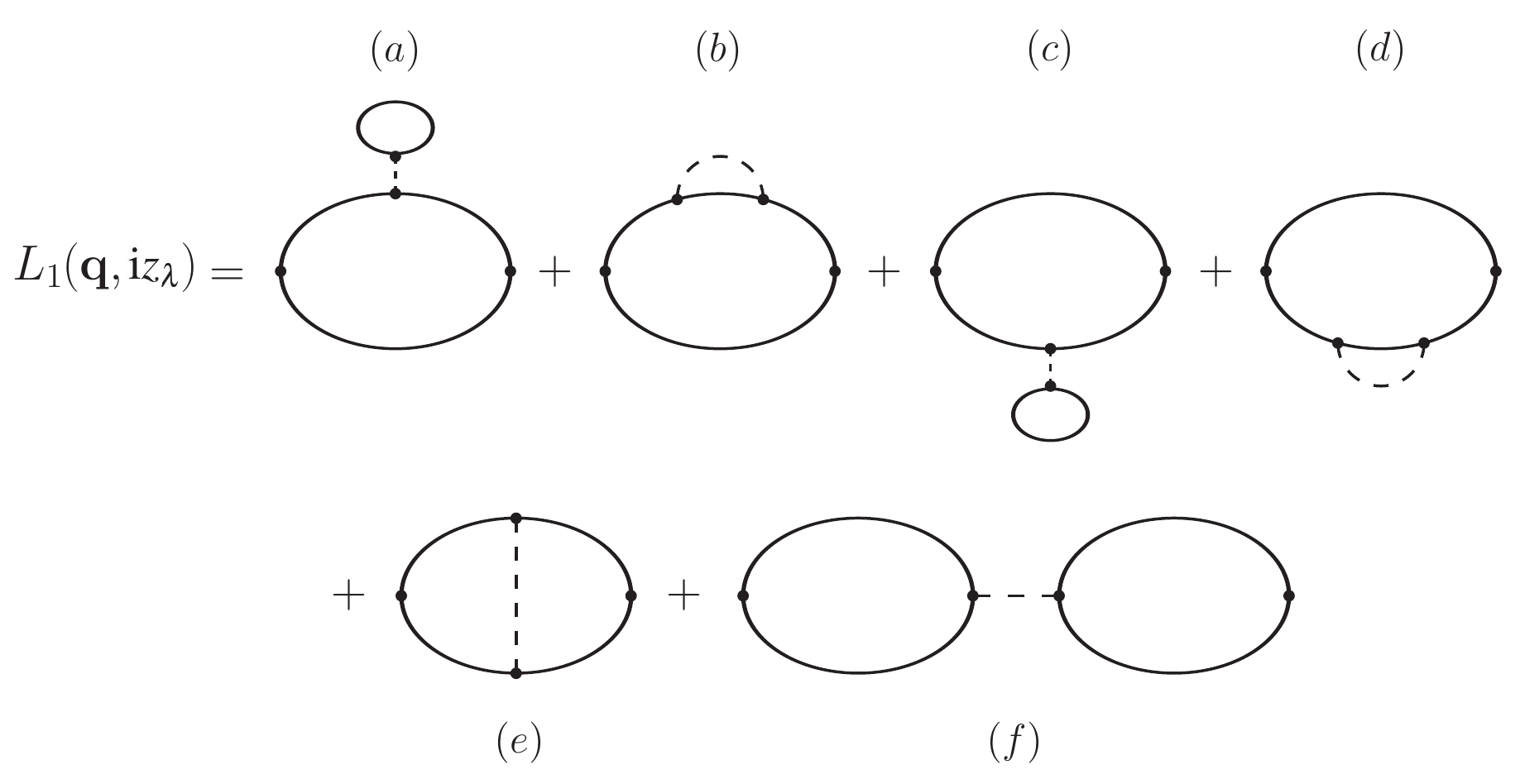}
\caption{The first-order perturbative contributions to the density auto-correlation function in Matsubara technique, $L_{1}({\bf q},\ii z_\lambda)$ (first order of interaction).}
\label{fig:1a}
\end{figure}
The contributions (a) -- (d) of Fig. \ref{fig:1a} are contributions to the self-energy (SE)
in Hartree-Fock approximation. They are taken into account if we replace the free
single-nucleon propagator by the quasiparticle propagator with the spectral function
$A^{\rm HF}(\omega , {\bf{p}})$, see Eq. (\ref{EHF}). The evaluation is similar
to the evaluation of $L_0$, Eq. (\ref{Lo}),
\begin{equation}\label{L1SE}
L_1^{\rm SE}({\bf q},z)=
g\int \frac{d^3 p}{(2\pi)^3}\frac{ f(\epsilon^{\rm HF}_p)-f(\epsilon^{\rm HF}_{{\bf p}+{\bf q}})
}{z+\epsilon^{\rm HF}_{p}-\epsilon^{\rm HF}_{{\bf p}+{\bf q}}}.
\end{equation}
The contribution (e) is a vertex correction,
and the exchange term (f) is a contribution to the screening equation.
The evaluation gives with quasiparticle propagators for the vertex (v) and screening (s) contribution
\begin{eqnarray}\label{L1v}
L_1^{\rm v,s}({\bf q},z)&=&
g\int \frac{d^3 p}{(2\pi)^3}\int \frac{d^3 k}{(2\pi)^3}\frac{ f(\epsilon^{\rm HF}_p)-f(\epsilon^{\rm HF}_{{\bf p}+{\bf q}})}{z+\epsilon^{\rm HF}_{p}-\epsilon^{\rm HF}_{{\bf p}+{\bf q}}}
\frac{ f(\epsilon^{\rm HF}_k)-f(\epsilon^{\rm HF}_{{\bf k}+{\bf q}})}{z+\epsilon^{\rm HF}_{k}-\epsilon^{\rm HF}_{{\bf k}+{\bf q}}}
\nonumber \\ &&
\times V({\bf p}+{\bf q},{\bf k};{\bf p},{\bf k}+{\bf q})_{\rm ex}.
\end{eqnarray}
Performing the limit $z \to \omega+ \ii 0$ we have
\begin{eqnarray}
S_1({\bf q},\omega)&=& \frac{1}{e^{\omega/T}-1}
g\int \frac{d^3 p}{(2\pi)^3}
\frac{\left( f(\epsilon^{\rm HF}_p)-f(\epsilon^{\rm HF}_{{\bf p}+{\bf q}})\right)
\left( f(\epsilon^{\rm HF}_k)-f(\epsilon^{\rm HF}_{{\bf k}+{\bf q}})\right)}{\epsilon^{\rm HF}_{p}-
\epsilon^{\rm HF}_{{\bf p}+{\bf q}}
+\epsilon^{\rm HF}_{k}-\epsilon^{\rm HF}_{{\bf k}+{\bf q}}}
\nonumber \\ &&
\times \left[\delta(\omega+
\epsilon^{\rm HF}_{p}-\epsilon^{\rm HF}_{{\bf p}+{\bf q}})+\delta(\omega+\epsilon^{\rm HF}_{k}-
\epsilon^{\rm HF}_{{\bf k}+{\bf q}}) \right] V({\bf p}+{\bf q},{\bf k};{\bf p},{\bf k}+{\bf q})_{\rm ex}.
\end{eqnarray}
After the limit ${\bf q} \to 0$ we arrive at the expression
\begin{eqnarray}
&&
\kappa^{(1)}_{\rm iso}(T,\mu)
=\frac{g}{n^2T}
\int \frac{d^3 p}{(2\pi)^3}f(\epsilon^{\rm HF}_p) [1-f(\epsilon^{\rm HF}_p)]\nonumber \\
&& \times \left[1-\frac{1}{T} \int \frac{d^3k }{(2 \pi)^3}V({\bf p},{\bf k};{\bf p},{\bf k})_{\rm ex}
f(\epsilon^{\rm HF}_k)[1-f(\epsilon^{\rm HF}_k)] \right]
\end{eqnarray}
in full agreement with Eq. (\ref{muHFn}) up to first order with respect to the interaction.

We conclude that we can reproduce the result (\ref{KTnB}) and the corresponding
relation (\ref{nHF}), starting from the density auto-correlation function or the dynamic
structure factor. As seen from the first-order calculation, the second way to
calculate the EoS using the dynamic structure factor is rather cumbersome compared with
the direct calculation of the density via the relation (\ref{neos}),
starting from the single-nucleon spectral function. However, the  dynamic structure factor contains
much more
information not only about thermodynamic properties, but also on dynamic properties and
the response to external perturbations. Before solving the question how bound state formation can be implemented within the diagram expansion of $L({\bf q},\omega)$, see Sec. \ref{two}, we present the Fermi-liquid approach which is a very efficient approach to nuclear systems.

\section{Fermi-liquid approach}\label{sec:Fermi}

One possible way to evaluate the dynamic structure factor is the Fermi-liquid approach as worked out by
Landau and Migdal. It is based on the  QPA using an effective  nucleon-nucleon interaction.
This concept has been applied successfully to describe also the dynamic behavior of dense matter.
We outline some important results here. However, until now there exists no systematic approach to include
the formation of bound states. In the following  Sec.  \ref{two} we show how the dynamic structure has to be treated
to include the formation of clusters.

In the low-density limit, we can infer an interaction potential $V_{\tau\tau^{'}}(1,2;1',2')$
which reproduces the two-particle properties such as scattering phase shifts
and bound state formation.
Different potentials such as the local Yukawa potential
or the separable Yamaguchi potential
are known from the literature.
For dense systems they must be modified to obtain the known properties at saturation density $n_{\rm sat}$.
Empirical interactions such as the Skyrme interaction can be introduced,
another possibility are relativistic mean-field (RMF) expressions derived
from an effective Lagrangian containing interacting nucleon and meson fields.

Dense nuclear systems at low temperatures are degenerate. Because of the Pauli blocking,
interaction processes are possible only near the Fermi surface at $p_{\rm F}$, and the potential $V_{\tau\tau^{'}}(1,2;1',2')$ is of relevance only for such processes. Therefore only the direction $\bf n$ of the wave number,
${\bf p}\approx p_{\rm F} \cdot {\bf n}$, is changing owing to the interaction.
The Landau Fermi-liquid approach considers such processes and introduces
corresponding phenomenological Landau-Migdal interaction parameters.
In this sense the Fermi-liquid approach is a semi-empirical approach.

Being treated within the Fermi-liquid theory the 4-point (like-sign, if treated in terms of non-equilibrium Green function technique) particle-hole interaction $T_{\rm ph}$ is presented as the local interaction  between particle-hole loops, cf. \cite{Voskresensky:1993ud},
\begin{eqnarray}
\label{vertices0}
\parbox{6cm}{\includegraphics[width=6cm]{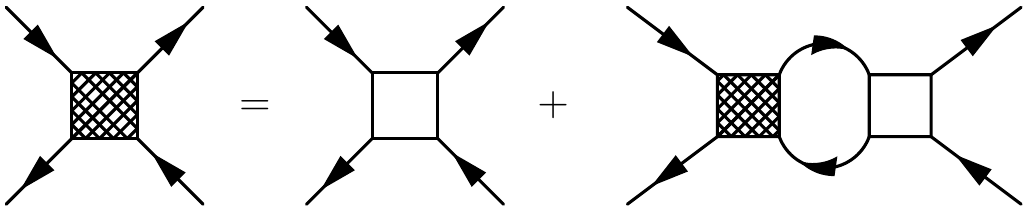}}
\end{eqnarray}
or
\begin{eqnarray}
\hat T_{\rm ph}({\bf n}',{\bf n};q)=\hat \Gamma^\omega({\bf n}',{\bf n})+\langle \hat \Gamma^\omega({\bf n}',{\bf n}'' {\cal L}_{\rm ph}({\bf n}'';q)\hat T_{\rm ph}({\bf n}'',{\bf n};q)\rangle_{{\bf n}''}.\nonumber
\end{eqnarray}
The single-particle Green functions are  taken in QPA,
the remaining background is encoded in the renormalized particle-hole interaction $\hat \Gamma^\omega({\bf n}',{\bf n})$.
The particle-hole propagator is given by the Lindhard function 
\begin{equation}
 {\cal L}_{\rm ph}({\bf n}'';q)=\int_{- \infty}^\infty \frac{d \epsilon}{2\pi i}\int_0^\infty \frac{dp \,\, p^2}{\pi^2} G(p_{\rm F +}) G(p_{\rm F -})\,,
\end{equation}
$p_{{\rm F}
\pm}=(\epsilon\pm\omega/2,p_{{\rm F}} \cdot{\bf{n}}\pm{\bf{k}}/2)$. The QPA and the improved quasiparticle approximation (IQPA) for the single-nucleon spectral function $A(p)$ are discussed in Appendix \ref{sec:quasi}.
The empty box, $\Gamma^{\omega}$, describes the $\delta$-functional interaction which  can be expressed through  the Landau-Migdal parameters as \cite{Mig,Mig1,Nozieres,GP-FL} 
\begin{align}
\hat\Gamma^{\omega}({\bf{n}}\,',{\bf{n}}) =
\Gamma^{\omega}_0({\bf{n}}\,'{\bf{n}})\,  \sigma'_0 \sigma_0 +
\Gamma^{\omega}_1({\bf{n}}\,'{\bf{n}})\,
(\vec{\sigma}' \cdot \vec{\sigma})\,.
\label{Gom-fullspin}
\end{align}
The matrices $\sigma_\mu$ with $\mu =0,\dots,3$ act on incoming
nucleons while the matrices $\sigma'_\mu$ act on outgoing nucleons;
$\sigma_0$ is the unity matrix and other Pauli matrices
$\sigma_{1,2,3}$ are normalized as ${\rm
Tr}\sigma_\mu\sigma_\nu=2\delta_{\mu\nu}$, and ${\bf{n}}\,',{\bf{n}}$ are unit vectors. We neglect here the spin-orbit
interaction, being suppressed for small transferred momenta
$|{\bf q}|\ll p_{\rm F}$ of our interest. The scalar and spin amplitudes in
Eq.~(\ref{Gom-fullspin}) can be expressed in terms of
dimensionless  scalar and spin Landau-Migdal parameters
\begin{align}
{f}_{\tau',\tau}({\bf{n}}\,',{\bf{n}})
&={\cal N}\Gamma_{0,\tau',\tau}^\omega({\bf{n}}\,',{\bf{n}})\,, \nonumber\\
{g}_{\tau',\tau}({\bf{n}}\,',{\bf{n}}) &={\cal N}
\Gamma_{1,\tau',\tau}^\omega({\bf{n}}\,',{\bf{n}})\,.
\label{L-param}
\end{align}
The normalization factor
\begin{equation}
 {\cal N}=g m_{\rm F}^*\,p_{\rm F}/(2\pi^2)
\end{equation}
contains the degeneracy factor  $g =2$ for one type of non-relativistic  fermions, like for the neutron matter,  and $g =4$ for two types of fermions, like for the isospin-symmetric nuclear matter, $\tau',\tau$ relate to $n$ or $p$.  In the case of a relativistic Fermi liquid one uses the normalization factor with $m^*_{\rm F}$ replaced by $E_{\rm F}=\sqrt{p_{\rm F}^2 +m^{*2}_{\rm F}}$. The baryon density, $n$, and the Fermi momentum, $p_{\rm F}$, are related as $n =g\,p_{\rm F}^3/(6\pi^2)$. Here we use normalization like in \cite{Nozieres,GP-FL}.
Note that another normalization on ${\cal N}(n_{\rm sat})$ instead of ${\cal N}(n)$ is used in \cite{Mig,Mig1}, where $n_{\rm sat}$ is the nuclear saturation density.

The generalization to a two-component system, e.g., to the nuclear
matter of arbitrary isotopic composition, is formally simple~\cite{Mig,Mig1}. Then, the amplitudes in ($nn$, $pp$, $np$ and $pn$) channels are in general different and equations for the partial amplitudes do not decouple. For the isospin-asymmetric case one also uses a symmetric normalization on $\sqrt{{\cal N}_{n} {\cal N}_{p}}$ with ${\cal N}_{\tau} =2 m_{\rm F}^*\,p_{\rm F,\tau}/\pi^2$, cf. \cite{Sawyer:1989nu}.

Bearing in mind the application to isospin-symmetric nuclear matter, we  present
$\Gamma_0^\omega,\,\Gamma_1^\omega$ in Eqs.~(\ref{Gom-fullspin}) and (\ref{L-param})  as ~\cite{Mig,Mig1}
\begin{eqnarray}
\Gamma_0^\omega =({f}({\bf{n}}\,',{\bf{n}})+{f}'({\bf{n}}\,',{\bf{n}})\vec{\tau}' \cdot \vec{\tau})/{\cal N}\,,\quad
\Gamma_1^\omega =({g}({\bf{n}}\,',{\bf{n}})+{g}'({\bf{n}}\,',{\bf{n}})\vec{\tau}' \cdot \vec{\tau})/{\cal N}\,,
 \end{eqnarray}
 where $\vec{\tau}$ are the isospin Pauli matrices. In this parametrization, the quantities $f$ and $f'$ (and similarly $g$ and $g'$) are expressed through $f_{nn}$ and $f_{np}$ as $f=\frac{1}{2}(f_{nn}+f_{np})$ and $f'=\frac{1}{2}(f_{nn}-f_{np})$. For the neutron matter  the parameters $f=f_{nn}$, $g=g_{nn}$ are the neutron-neutron Landau-Migdal scalar and spin parameters. For isospin-symmetric nuclear  systems we have  $f_{nn}=f_{pp}$ and $f_{np}=f_{pn}$ if Coulomb interaction is omitted.

The particle-hole interaction is adapted to known properties of nuclear matter so that it is parametrized in an empirical way. This can be done directly comparing with known properties near the saturation density $n_{\rm sat}$
of baryons.
The  parameters $f({\bf{n}}\,',{\bf{n}})$ and $g({\bf{n}}\,',{\bf{n}})$ depend on the scattering angle and are expanded in  Legendre polynomials,
\begin{eqnarray}
f_{\tau',\tau}({\bf{n}},{\bf{n}}')=\sum_l f_{l,\tau',\tau} P_l ({\bf{n}},{\bf{n}}')\,, \quad g_{\tau',\tau}({\bf{n}},{\bf{n}}')=\sum_l g_{l,\tau',\tau} P_l ({\bf{n}},{\bf{n}}')\,.
 \end{eqnarray}
For the most important physical quantities only  the $l=0,1$ harmonics contribute.
For instance, the incompressibility $K$ is related to the Landau-Migdal parameter $f_0$ as
\begin{equation}
 K=n\frac{\partial \mu}{\partial n}{\Big |}_T =\frac{p_{\rm F}^2}{3E_{\rm F}} (1+f_0).
\end{equation}
 The Landau effective mass of the nucleon quasiparticle can be expressed in terms of the Landau-Migdal parameter $f_1$,  \begin{equation}
 E_{\rm F}=\mu (1+f_1/3)\,.
 \end{equation}
(Note that in the relativistic theory the particle energy at the Fermi surface plays the same role as the effective mass in the non-relativistic Fermi-liquid theory.)

The Landau-Migdal parameters can be extracted from the experimental data on atomic nuclei. Unfortunately, there are essential uncertainties in numerical values of some of
these parameters. These uncertainties are, mainly, due to attempts to get the best fit
to experimental data in each concrete case slightly modifying parametrization used
for the residual part of the $N-N$ interaction.
For example, basing on the analysis of Refs.~\cite{KS80}, with the normalization \cite{MSTV90} $C_0 =1/ N_0 =300$\,MeV$\cdot$fm$^{3}$  one gets ${f}_0 \simeq 0.25$, ${f}_0' \simeq 0.95$, ${g}_0 \simeq 0.5$, ${g}_0' \simeq 1.0$, see Table 3 in~\cite{Mig1}. The parameters ${g}_0 $, ${g}_0'$   are rather slightly density dependent whereas ${f}_0 $ and   ${f}_0'$ depend on the density essentially.  For ${f}_0 (n)$ Ref.~\cite{Mig} suggested to use a  linear density dependence, then with above given parameters we have  ${f}_0 (n)=-2.5+2.75\, n/n_{\rm sat}$.  For strongly isospin-asymmetric matter, e.g. for neutron matter, there are no data from which the Landau-Migdal parameters can be extracted. In this case the parameters are calculated within a chosen model for the $N-N$ interaction, see the review \cite{Backman:1984sx}.

We are interested in the description of the scalar interaction channel. Then  the empty box is
\begin{eqnarray}
\Gamma^\omega \equiv F (\theta)\,,\,\,\, \theta ={\bf{p}}\cdot {\bf{p}}{\,'}/ |{\bf{p}}||{\bf{p}}{\,'}|\,.
\end{eqnarray}
Simplifying our considerations we will retain only the zeroth harmonics $F_0$ and $f_0$.
For isospin-symmetric matter we take the combinations  $F_0 =(F_{0,nn}+F_{0,np})/2$ and $f_0=(f_{0,nn}+f_{0,np})/2$  but for pure neutron matter $ F_{0,nn}$ and $f_{0,nn}$. Then (\ref{vertices0}) produces
\begin{eqnarray}
T^{\rm ph}_{- -}=F_0/[1+F_0 \Pi_{0,- -}^{N=1,00}]\,,
\end{eqnarray}
$\Pi_{0,- -}^{N=1,00}$ corresponds to the first one $- +$ loop term in $(G^{-+}_{\rm F}G^{+-}_{\rm F})$ loop expansion, cf.   \cite{Knoll:1995nz} and Appendix \ref{sec:spectral}. In the standard Fermi-liquid approach for equilibrium systems,  Eq. (\ref{vertices0}) is treated as equation for the retarded quantities \cite{Voskresensky:1993ud}. Resummation yields
$$T^{R}_{\rm ph} =F_0/[1+F_0 \Pi_{0}^{R}].$$

With the  Fermi-liquid interaction follows
\begin{eqnarray}\label{Reself}
\mbox{Re}\Pi^R (0,{\bf{q}}\to 0)=\mbox{Re}\Pi^R_0 (0,{\bf{q}}\to 0)/[1+F_0\mbox{Re}\Pi^R_0 (0,{\bf{q}}\to 0)]\,.
\end{eqnarray}
Thus from (\ref{0EqstSpectr}) and (\ref{Reself}) we derive
\begin{eqnarray}\label{munF}
S({\bf{q}}\to 0)=
\frac{T}{\partial \mu/\partial n}=
\frac{T\,\mbox{Re}\Pi^R_0 (0,{\bf{q}}\to 0)}{[1+F_0\mbox{Re}\Pi^R_0 (0,{\bf{q}}\to 0)]}\,.
\end{eqnarray}
For $T\to 0$ one has $\mbox{Re}\Pi^R_0 (0,{\bf{q}}\to 0)={\cal N}$ and using (\ref{EqStstr0}) we derive ordinary relation for the Fermi liquid
\begin{equation}
 \frac{\partial \mu}{\partial n}=\frac{1}{{\cal N}}+F_0 =\frac{1+f_0}{{\cal N}}.
\end{equation}

\section{Density-density correlations within relativistic mean-field approximation}\label{RMF}
\subsection{Self--consistent Hartree approximation}

To simplify our considerations we continue the study of pure neutron or isospin-symmetric matter.
The QPA and IQPA are discussed in the Appendix \ref{sec:spectral}.
Within the self-consistent Hartree approximation $\Sigma$ depends on $n$ and $\mu$  via the dependence of the Dirac effective fermion mass $m^* (n)$
and $f$ on $\mu -V_0$. Then Eq. (\ref{dmun}) being treated  within the IQPA becomes
\begin{eqnarray}\label{dmunH}
\frac{\partial \mu}{\partial n}-\frac{\partial V_0}{\partial n}=\frac{T+ h_s^{\rm IQPA} (\partial m^* /\partial n)}{ h^{\rm IQPA}}\,,
\end{eqnarray}
\begin{eqnarray}
h^{\rm IQPA}=g\int\frac{\di^3 p}{(2\pi)^3}f(E_p -\mu +V_0)(1-f(E_p -\mu +V_0))\,,
\end{eqnarray}
\begin{eqnarray}\label{hsiqpa}
h^{\rm IQPA}_s=g\int\frac{\di^3 p}{(2\pi)^3}  \frac{m^*}{E_p}f(E_p -\mu +V_0)(1-f(E_p -\mu +V_0)).
\end{eqnarray}
With $m^*$ not depending on $\mu$ and taking $V_0 =F_0 n$ we recover (\ref{munF}), now in the IQPA.

Reference \cite{Matsui} calculated the Landau-Migdal parameter $f_0$ in the  non-linear Walecka model
with the RMF Lagrangian for the nucleons interacting  with $\omega_0$ and $\sigma$ mean fields of $\omega$ and $\sigma$ mesons.
The generalization to include the $\rho_0^3$ meson field of the $\rho$ meson  is straightforward. In this approach the nucleon distribution is given by
\begin{eqnarray}
f =\frac{1}{\mbox{exp}[(E_p -\mu+V_0)/T]+1}\,,\quad E_p=\sqrt{m^{\,*2}+{\bf{p}}^{\,2}}\,,\quad V_0 =\frac{g^2_\omega n}{m^2_\omega}\,,
\end{eqnarray}
$g_{\omega }$ is the $\omega N$ coupling, $m_\omega$ is the $\omega$ meson mass,
and the Dirac effective nucleon mass obeys the  equation
\begin{eqnarray}\label{efmass}
m^* =m-\frac{g^2_\sigma}{m^2_\sigma}\int \frac{g\di^3 p}{(2\pi)^3}f \frac{m^*}{E_p}\,,
\end{eqnarray}
$g_{\sigma}$ is the $\sigma N$ coupling, $m_\sigma$ is the $\sigma$ meson mass.
Taking $\partial/\partial n$ in (\ref{efmass}) we find
\begin{eqnarray}
\frac{\partial m^*}{\partial n}=-\frac{g^2_\sigma}{m^2_\sigma}\left(\frac{\partial \mu}{\partial n}- \frac{\partial V_0}{\partial n}\right)\frac{m^*  h_s}{T+\frac{g^2_\sigma}{m^2_\sigma} h_{s1}}\,,
\end{eqnarray}
the integral $ h_{s1}/T$ is reduced to
\begin{eqnarray}
\frac{ h_s}{T}=g \int_{m^*}^{\infty}\frac{ E_p \di E_p f}{2\pi^2 \sqrt{E_p^2 -m^{\,*2}}}\,,
\end{eqnarray}
\begin{eqnarray}
\frac{h_{s1}}{T}= g \int_{m^*}^{\infty}\frac{ (E_p^2 -2m^{\,*\,2}) \di E_p f}{2\pi^2 \sqrt{E_p^2 -m^{\,*2}}}\,,
\end{eqnarray}
and
\begin{eqnarray}
\frac{ h}{T} =g \int_{m^*}^{\infty}\frac{(2E_p^2 -m^{\,*\,2}) \di E_p f}{2\pi^2 \sqrt{E_p^2 -m^{\,*2}}}\,.
\end{eqnarray}
With these quantities at hand using (\ref{hsiqpa}) we find
 \begin{eqnarray}\label{dmunHMatsui}
 S({\bf{q}}\to 0)=\frac{T}{\partial \mu/\partial n}=\frac{h}{1+F_0^T h/T}\,,
\end{eqnarray}
where the quantity $F_0^T$ has the meaning of the zero harmonic of the scalar Landau-Migdal parameter for $T\neq 0$,
\begin{eqnarray}\label{MatsuiT}
F_0^T = \frac{g^2_\omega }{m^2_\omega}-\frac{g^2_\sigma }{m^2_\sigma}\frac{m^{\,*\,2} h_{s}^2}{h^2 [1+\frac{g^2_\sigma }{m^2_\sigma}(\   h_{s1}/T +m^{\,*\,2} h_s^2/hT)]}\,.
\end{eqnarray}
The latter result generalizes the result of Matsui \cite{Matsui}  for $T\neq 0$.
For $T=0$
\begin{eqnarray}
& h_s/T =\frac{g p_{\rm F}}{2\pi^2}\,,\quad h/T= \frac{g E_{\rm F} p_{\rm F}}{2\pi^2}\,\quad E_{\rm F}=\sqrt{p_{\rm F}^2 +m^{*\,2}}\,,\\
& h_{s1}/T =-\frac{p_{\rm F}^3}{E_{\rm F}}+\frac{3}{2} E_{\rm F} p_{\rm F}-\frac{3}{2}m^{*\,2}\ln \frac{E_{\rm F}+p_{\rm F}}{m^*}\,,\nonumber
\end{eqnarray}
and the result (\ref{MatsuiT})  coincides with that of \cite{Matsui}. (Note that for  $f_0$ and $F_0$ we use notations different from \cite{Matsui}.)
With $\partial \mu/\partial n$ at hand we may reconstruct the EoS, e.g. Eq.~(\ref{pressure}) for the pressure.

\subsection{Generalizations to isospin-asymmetric system}\label{arbitrary}
For a multi-component system, we have to generalize the expressions given in  Sec. \ref{FDT} introducing partial structure factors for the constituents. We use the relation, cf. \cite{Sawyer:1989nu},
\begin{equation}
\label{partSF}
\langle\rho_\tau(0)\rho_{\tau'}(0)\rangle-\langle\rho_\tau (0)\rangle\langle\rho_{\tau'}(0)\rangle =T\frac{\partial \langle\rho_\tau (0)\rangle }{\partial \mu_{\tau'}}
=TV\frac{\partial n_\tau }{\partial \mu_{\tau'}}
\end{equation}
where $\rho_\tau (q \to 0)=N_\tau$ is the particle number
according to Eq. (\ref{densityrhoq}).
Introducing
\begin{eqnarray}
&S_{nn}({\bf q}\to 0)=\langle\rho^{\dagger}_n({\bf{q}})\rho_n({\bf{q}})\rangle_{{\bf{q}\to 0}}/n_n\,,\quad S_{np}({\bf q}\to 0)=\langle\rho^{\dagger}_n({\bf{q}})\rho_p({\bf{q}})\rangle_{{\bf{q}\to 0}}/\sqrt{n_n n_p}\,,\\
&S_{pp}({\bf q}\to 0)=\langle\rho^{\dagger}_p ({\bf{q}})\rho_p({\bf{q}})\rangle_{{\bf{q}\to 0}}/n_p\,,\quad S_{pn}({\bf q}\to 0)=\langle\rho^{\dagger}_p({\bf{q}})\rho_n({\bf{q}})\rangle_{{\bf{q}\to 0}}/\sqrt{n_p n_n}\,,
\end{eqnarray}
the static structure factor can be presented as \cite{Burrows:1998cg}
\begin{equation}
S({\bf q}\to 0)=S_{nn}({\bf q}\to 0)+S_{pp}({\bf q}\to 0)+2S_{np}({\bf q}\to 0)\,,
\end{equation}
provided that $S_{np}=S_{pn}$.
Thus
\begin{equation}\label{Stf}
S_{\tau,\tau'}({\bf q}\to 0)=T\left(\frac{\partial n_\tau}{\partial \mu_{\tau'}}\right){\Big |}_T
\end{equation}
 and
\begin{equation}
\kappa_{\rm iso}(T,\mu)=
\frac{1}{n^2} \left[\left(\frac{\partial n_n}{\partial \mu_n}\right){\Big |}_T+ 2\left(\frac{\partial n_n}{\partial \mu_p}\right){\Big |}_T+\left(\frac{\partial n_p}{\partial \mu_p}\right){\Big |}_T\right]\,,
\end{equation}
 provided $\left(\frac{\partial n_n}{\partial \mu_p}\right){\big |}_T=
\left(\frac{\partial n_p}{\partial \mu_n}\right){\big |}_T$.

Within the RMF approach one may find the values of the  Landau-Migdal parameters in the scalar channel in a wide region of density, temperature, and asymmetry. For example
we can consider the RMF approach which claims to reproduce the known empirical data
and extrapolates for a wide range of temperatures and densities. Here
\begin{equation}
\label{RMF1}
n_\tau^{\rm RMF}(T,\mu_n,\mu_p)= \frac{1}{V} \sum_pf[E_\tau^{\rm RMF}({\bf p})-\mu_\tau]\,
\end{equation}
where the relativistic chemical potential $\mu_\tau$ includes the rest mass. The relativistic quasiparticle energies are given as
\begin{equation}
\label{RMF2}
E_\tau^{\rm RMF}({\bf p};T,n_n,n_p)=\sqrt{[m_\tau -S (T,n_n, n_p)]^2+{\bf p}^2}+V_\tau (T,n_n,n_p)
\,.
\end{equation}

\subsection{Landau-Migdal parameters and DD2-RMF model}
Values for $S(T,n,Y_p)$ and $V_\tau(T,n,Y_p)$ according to the Typel DD2-RMF EoS \cite{Typel,Typel1999} are given in Ref. \cite{R}.

{}From $\delta E_\tau =\sum_{\tau'} F_{\tau \tau'} \delta n_{\tau'}$ we find
$F^{\tau \tau'}_0 =(\partial E_\tau/\partial n_{\tau'})_{p_\tau=p_{{\rm F},\tau}}$.
So we have
\begin{equation}
F_0^{nn}=-\frac{(m_n-S)(\partial S/\partial n_n)}{\sqrt{[m_n -S (T,n_n, n_p)]^2+ p_{{\rm F},n}^2}}+\frac{\partial V_n}{\partial n_n}\,,
\end{equation}
\begin{equation}
F_0^{pp}=-\frac{(m_p-S)(\partial S/\partial n_p)}{\sqrt{[m_p -S (T,n_n, n_p)]^2+ p_{{\rm F},p}^2}}+\frac{\partial V_p}{\partial n_p}\,,
\end{equation}
\begin{equation}
F_0^{np}=-\frac{(m_n-S)(\partial S/\partial n_p)}{\sqrt{[m_n -S (T,n_n, n_p)]^2+ p_{{\rm F},n}^2}}+\frac{\partial V_n}{\partial n_p}\,,
\end{equation}
\begin{equation}
F_0^{pn}=-\frac{(m_p-S)(\partial S/\partial n_n)}{\sqrt{[m_p -S (T,n_n, n_p)]^2+ p_{{\rm F},p}^2}}+\frac{\partial V_p}{\partial n_n}\,,
\end{equation}
\begin{figure}[!th]
\includegraphics[width=10cm]{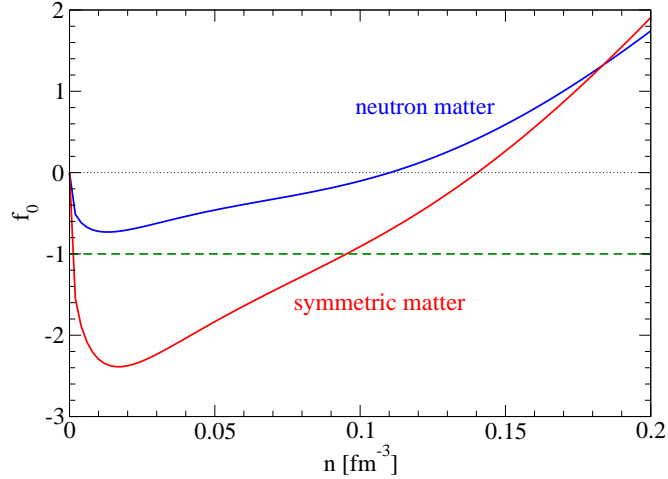}
\caption{Landau parameter $f_0(n)$ at $T=0$ for neutron matter ($Y_p=0$) and symmetric matter ($Y_p=0.5$), derived from the DD2-RMF approximation,  see Appendix \ref{DD2}.}
\label{fig:0}
\end{figure}

Fig. \ref{fig:0} shows the Landau-Migdal parameter $f_0(n,Y_p)$.
For symmetric matter, $f_0(n,0.5)$ a region of instability ($f_0(n)\le -1$) occurs
between density values $n=  0.000502$ and      $ 0.09528$ fm$^{-3}$ (spinodal instability).
For neutron matter  $f_0(n,0)$, no thermodynamic instability occurs. For details  see Appendix \ref{DD2}.
A more general discussion how to relate RMF and  Fermi-liquid theories is given in Ref. \cite{Matsui}.

\section{Two-particle correlations}\label{two}

\subsection{Cluster decomposition and inclusion of bound states}

We investigated the dynamic structure factor as a fundamental property
of the many-nucleon system which allows also to derive the thermodynamic properties,
in particular the EoS. We gave an extended discussion of the mean-field approximation
and showed the relation to the Landau-Migdal  Fermi-liquid theory. These semiempirical
approaches, based on some parameter values to characterize the interaction,
are very efficient to describe nuclear systems, at least, near the saturation density and at low excitation energies.
At low temperatures the system is degenerate, and part of correlations are suppressed
because of Pauli blocking. The quasiparticle concept is appropriate to describe excitations
at these conditions.

A problem arises when we are going to low density matter, because the strong interaction
between the nucleons is no longer blocked out. Below the so-called Mott density,
bound states can be formed. Properties of nuclear systems are significantly influenced
by these few-body correlations. However, as a quantum effect, bound states are
not easily included in a mean-field approach which works with single-particle properties.
Similarly, within the Thomas-Fermi or the Fermi-liquid model the incorporation of
bound-state formation is difficult, see Ref. \cite{KV16}. In this work, we outline
the method how to include bound-state formation into the theory of the dynamic structure factor
for nuclear systems.

\begin{figure}[!th]
\includegraphics[width=11cm]{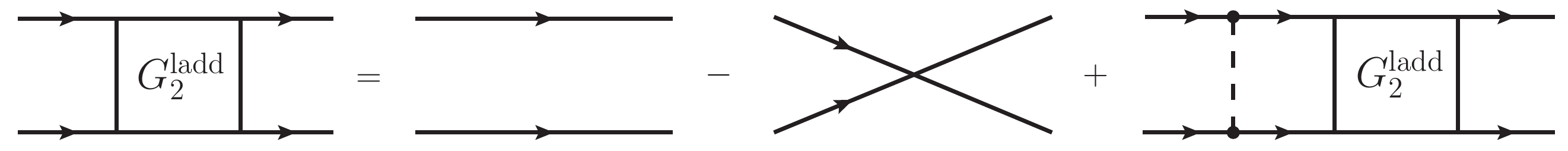}
\caption{Two-particle Green function in ladder approximation. }
\label{fig:G2}
\end{figure}

The QS approach to nuclear systems cannot describe bound state formation
in any finite order of perturbation theory but only after summation of infinite
orders of contributions. In the present section, we investigate the formation of clusters in warm nuclear matter in the
low-density limit,  for $n/\Lambda^3 \ll1, \Lambda^2 =2\pi/mT$ in the ladder approximation where we first neglect  in-medium effects on the single-particle Green functions. 
In Fig.  \ref{fig:G2} we present the two-particle Green
function such a ladder approximation. The QPA including medium effects will be discussed
in Sec. \ref{comp}.
We have with $G^{(0)}_1(1,\ii z_\nu)=1/(\ii  z_\nu- \epsilon_1)$
\begin{equation}
\label{ladd}
 G_2^{\rm ladd}(1,2;1',2',\ii z_\mu)=\sum_\nu G^{(0)}_1(1,\ii z_\nu) G^{(0)}_1(2,\ii z_\mu-\ii z_\nu)
\left[\delta_{1,1'}\delta_{2,2'}-\delta_{1,2'}\delta_{2,1'}+\sum_{34}V(1,2;3,4)
G_2^{\rm ladd}(3,4;1',2', \ii z_\mu)\right]
\end{equation}
The ladder sum leads to a  Schr\"odinger equation which describes
the quantum two-body problem. This so-called chemical picture
introduces bound states as new quasiparticles in a systematic way,
taking the corresponding sums of ladder diagrams into account.
In the low-density limit where medium effects  on the single-particle Green functions  are neglected,
we have in the representation with respect to the eigenstates of the two-nucleon system
$ \psi_{\nu,{\bf P}}({\bf p}_1,{\bf p}_2) $ with the energy
eigenvalues $ E^0_{\nu,{\bf P}} $ that can both be obtained as a solution of the free two-nucleon Schr{\"o}dinger equation
\begin{equation}
 \left( \frac{p_1^2}{2m}+\frac{p_2^2}{2m}-E^0_{\nu,{\bf P}}\right) \psi_{\nu,{\bf P}}({\bf p}_1,{\bf p}_2)+\sum_{p_1',p_2'} V({\bf p}_1,{\bf p}_2;{\bf p}'_1,{\bf p}'_2) \psi_{\nu,{\bf P}}({\bf p}_1',{\bf p}_2')=0,
\end{equation}
the expression
\begin{equation}
\label{G2op}
 G_2^{\rm ladd}(1,2;1',2', \ii z_\mu)=\sum_{\nu.{\bf P}} \psi_{\nu,{\bf P}}({\bf p}_1,{\bf p}_2)\frac{1}{\ii z_\mu -E^0_{\nu,{\bf P}}}
\psi^*_{\nu,{\bf P}}({\bf p}'_1,{\bf p}'_2),
\end{equation}
where ${\bf P}$ denotes the c.m. momentum and $\nu$ the intrinsic state of the two-body system.
The proof is easily given by insertion in Eq. (\ref{ladd}). Medium effects  in the  single-particle channel  are included considering
the single-nucleon propagator $ G_1(1,\ii z_\nu) $ in QPA and taking Pauli blocking
terms into account, see Refs. \cite{RMS,R,SRS} where an in-medium Schr{\"o}dinger equation is derived.
Our prescription to include cluster formation (bound states) into a systematic many-body approach
is to consider ladder propagators in addition to single-nucleon propagators (chemical picture).
Within a consistent approach, double counting of diagrams has to be avoided.

\subsection{The Beth-Uhlenbeck formula}
\label{sec:BU}

\begin{figure}[!th]
\includegraphics[width=10cm]{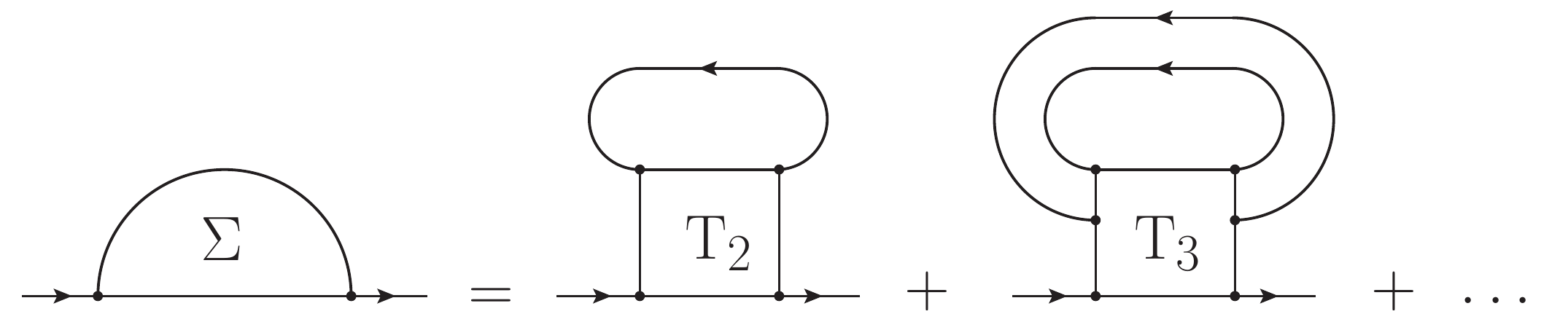}
\caption{Cluster decomposition of the single-nucleon self-energy.}
\label{fig:SE}
\end{figure}

In this section we consider the low density limit of the nuclear matter EoS,  assuming $n/\Lambda^3 \ll1, \Lambda^2 =2\pi/mT$ and using the ladder approximation where the Beth-Uhlenbeck formula is obtained for the second virial coefficient. To include cluster formation in the EoS, we consider firstly the density EoS (\ref{neos})  with the  single-particle spectral function obtained from the cluster  decomposition of the self-energy shown in Fig. \ref{fig:SE}, see Refs. \cite{RMS,SRS}. To calculate the nucleon self-energy  $\Sigma(1,z)$ we introduce the few-particle $T_A$ matrices describing the
 $A$ - nucleon cluster in the low-density limit.

In particular we have
\begin{equation}
 T_2^{\rm ladd}(1,2;1',2', \ii z_\mu)=V(1,2;1'',2'') G_2^{\rm ladd}(1'',2'';1',2', \ii z_\mu)[ G_{2}^{0}(1',2', \ii z_\mu)]^{-1}\,
\end{equation}
with $G_{2}^{0}$ being the contribution to $G_2^{\rm ladd}$ in zeroth order of interaction. The $T_A$-matrix is the amputated part of the $G_A$ function, the propagator
of the $A$-nucleon cluster, in analogous manner to $G_2$.
The evaluation of the  self-energy is straightforward,  see \cite{RMS,SRS}.
As shown there,  the spectral function is obtained from the imaginary part of the self-energy, and the density in the Beth-Uhlenbeck (BU) approach
follows according (\ref{neos}), (\ref{EQdistr}) as 
\begin{equation}
\label{BU0}
n^{\rm BU}(T, \mu)=\frac{1}{V} \sum_{\bf p}f(\epsilon_p^0)+\frac{2}{V} \sum_{\nu,{\bf P}}
f_{\rm B}(\epsilon^0_{\nu,{\bf P}})-\frac{2}{V} \sum_{{\bf p}_1,{\bf p}_2}f_{\rm B}(\epsilon_{p_1}^0+\epsilon_{p_2}^0).
\end{equation} 
Here, $f_{\rm B}(z)=[\exp(z/T)-1]^{-1}$ denotes the Bose distribution, with the free two-particle energies $\epsilon^0_{\nu,{\bf P}}=E^0_{\nu,{\bf P}}-\mu_1-\mu_2$. As already said, $\nu$ describes the intrinsic state of the two-body system, in particular bound and scattering states. The last term in (\ref{BU0}) describes the free particle contribution contained in the scattering states which must be subtracted.
Considering only the bound state part of the sum over the intrinsic quantum number $\nu$, see Eq. (\ref{G2op}),
the mass action law (nuclear statistical equilibrium, NSE) is reproduced for the two-nucleon bound-state formation.

In contrast to the NSE, the sum over the intrinsic quantum number $\nu$ in (\ref{BU0}) includes also the scattering states.
This is inevitable to obtain the correct second virial coefficient, which determines the second order of
density in the EoS $\mu(T,n)$. For the scattering states, we introduce, in addition to the intrinsic spin state $\alpha$,
the relative momentum ${\bf p}_{\rm rel}$
instead of the intrinsic quantum number $\nu$. Similarly, ${\bf p}_1,{\bf p}_2$ is
replaced by ${\bf P}, {\bf p}_{\rm rel}$ in the last contribution of Eq. (\ref{BU0}).
We transform to the energy $E_{\rm rel}=p_{\rm rel}^2/m$ and replace the summation over ${\bf p}_{\rm rel}$
by the integral over $E_{\rm rel}$ where the density of states $D(E_{\rm rel})$ is introduced.
For the free two-nucleon propagator,
the density of states is simply given by $D^0(E_{\rm rel})=Vm \,p_{\rm rel}/(4 \pi^2)$.
For the interacting two-particle system,
the density of states is related to the scattering phase shift $\delta_{\alpha, {\bf P}}(E_{\rm rel})$
in the channel $\alpha$ (describing the spin/isospin state),  see \cite{LL1980}, Sect. VII, so that we have for the scattering part
\begin{equation}
 D^{\rm scat}_{\alpha, {\bf P}}(E_{\rm rel})=g_\alpha \left(V  \frac{m \,p_{\rm rel}}{4 \pi^2}+\frac{\partial}{\partial E_{\rm rel}} \delta_{\alpha, {\bf P}}(E_{\rm rel}) \right),
\end{equation} 
where $g_\alpha$ denotes the degeneracy factor such as $g_\alpha=3$ for the spin-triplet channel where the deuteron is formed.

 Altogether, from the QS approach we obtain the
Beth-Uhlenbeck formula \cite{SRS} (the first part of $D^{\rm scat}_{\alpha,{\bf P}}(E_{\rm rel})$ is compensated by the free nucleon contribution $D^0(E_{\rm rel}) $),
\begin{equation}
\label{BU}
n^{\rm BU}(T, \mu)=\frac{1}{V} \sum_{\bf p}f(\epsilon_p^0)+\frac{2}{V} \sum_{\alpha,{\bf P}}\int_{-\infty}^\infty \frac{dE_{\rm rel}}{\pi}
f_{\rm B}\left(E_{\rm rel}+\frac{P^2}{4m}-\mu_1-\mu_2 \right) D^{\rm BU}_{\alpha, {\bf P}}(E_{\rm rel}),
\end{equation}
where we implemented the contribution of the bound states (second term of the right-hand side of (\ref{BU0}) so that
\begin{equation}
\label{DBU}
 D^{\rm BU}_{\alpha, {\bf P}}(E_{\rm rel})=g_\alpha \left(\sum_{\nu'} \pi \delta(E_{\rm rel}-E^0_{\alpha \nu', {\bf P}})
 +\frac{\partial}{\partial E_{\rm rel}} \delta_{\alpha, {\bf P}}(E_{\rm rel}) \right)
\end{equation}
is the density of states which contains the  bound state energy
$E^0_{\alpha \nu', {\bf P}}=E^0_{\alpha \nu'}+{\bf P}^2/4m$  and scattering phase shift $\delta_{\alpha, {\bf P}}(E)$ as function of the
 energy $E$ of relative motion. For arbitrary mass numbers $A$, the intrinsic quantum number $\nu'$ denotes ground as well as possible excited bound sates in the spin/isospin channel $\alpha$.
Instead of a separate sum over bound states, cf. Eq. (\ref{BU0}),
we included the bound state contribution as $ \pi \delta(E-E^0_{\alpha \nu', {\bf P}})$ in the density of states,
contributing at negative values of the general variable $E_{\rm rel}$. 

Note that for Breit-Wigner resonances, see \cite{LL3}, Chapter XVII,
\begin{eqnarray}
\mbox{tan} \delta_{\rm R} =-\frac{\Gamma_{\rm R}}{2M}\,,\quad {\rm{and}}\quad  \frac{\partial \delta_{\rm R}}{\partial E} = \frac{A}{2} =\frac{\Gamma_{\rm R}/2}{M^2 +\Gamma_{\rm R}^2/4}\,,
\end{eqnarray}
where $M=E-E_{\rm R}\,,$ $E=E_{\rm R}-\ii \Gamma_{\rm R}/2\,,$
 cf. Appendix \ref{A1}, Eq. (\ref{eqrel1}). More general relations can be found in
\cite{Kolomeitsev:2013du}.

The Beth-Uhlenbeck formula is an exact expression for the second virial coefficient $b$ defined by
the virial expansion of the EoS \cite{BU,R,SRS} (valid for $n_\tau \Lambda^3 \ll 1$). At given $T$, the pressure is considered as function of the density, and the prefactors of a power expansion are denoted as virial coefficients $b$. This power expansion  can also be done for the chemical potential, and the inversion gives the relations
\begin{eqnarray}\label{bnnp}
&&n_n(T,\mu_n,\mu_p)=\frac{2}{\Lambda^3} \left[b_n(T) e^{\mu_n/T}+2b_{nn}(T)e^{2 \mu_n/T}+2 b_{np}(T) e^{(\mu_n+\mu_p)/T}
+\dots \right],
\nonumber \\
&&n_p(T,\mu_n,\mu_p)=\frac{2}{\Lambda^3} \left[b_p(T) e^{\mu_p/T}+2b_{pp}(T)e^{2 \mu_p/T}+2 b_{pn}(T) e^{(\mu_n+\mu_p)/T}
+\dots \right],
\end{eqnarray}
where $ \Lambda^2=2 \pi /m T $.  As well known, in the low-density limit all systems behave at finite $T$ like ideal classical gases so that $b_n(T)=b_p(T)=1$.
The second virial coefficient contains the effects of degeneration as well as interaction terms.
In particular, we have after performing integration by parts and using the Levinson theorem
\begin{equation}\label{bnn}
 b_{nn}(T)=-\frac{1}{2^{5/2}}+ \frac{1}{2^{1/2} \pi T} \int_0^\infty dE e^{-E/(2T)} \delta_{nn}(E)
\end{equation}
and \begin{equation}\label{bnp}
 b_{np}(T)=\frac{3}{2^{1/2}} [e^{E^0_d/T}-1]+ \frac{1}{2^{3/2} \pi T} \int_0^\infty dE e^{-E/(2T)} \delta_{np}(E)\,.
\end{equation}
The deuteron (binding energy $E^0_d=2.225$ MeV) arises in the isospin-singlet, spin-triplet channel so that the degeneracy factor is $g_d=3$. The scattering phase shifts $\delta_{nn}(E),\delta_{np}(E)$ are given by the contributions of the different channels, see also \cite{R,SRS,HS} for details. Assuming symmetric matter we have $b_{pp}=b_{nn},b_{pn}=b_{np} $  provided Coulomb effects are disregarded.

From (\ref{Stf}) and (\ref{bnnp}) we find relations between the partial static structure factors and the virial coefficients (\ref{bnn}), (\ref{bnp}):
\begin{eqnarray}
&&S_{nn}(T,\mu_n,\mu_p)=
\frac{2}{\Lambda^3}\left[b_n(T) e^{\mu_n/T}+4b_{nn}(T)e^{2 \mu_n/T}+2 b_{np}(T) e^{(\mu_n+\mu_p)/T}
+\dots \right]\,, \\
&&S_{np}(T,\mu_n,\mu_p)=\frac{4}{\Lambda^3}b_{pn}e^{(\mu_n+\mu_p)/T}+...\,,\nonumber\\
&&S_{pp}(T,\mu_n,\mu_p)=
\frac{2}{\Lambda^3}\left[b_{p}(T)e^{\mu_p/T}+4b_{pp}(T)e^{2\mu_p/T}+
2b_{pn}(T)e^{(\mu_n+\mu_p)/T}+\dots \right]\,.\nonumber
\end{eqnarray}
One can also express $S_{nn}$, $S_{np}$, $S_{pp}$ in terms of $n_n$ and $n_p$ variables. Simple explicit expressions are obtained for the second order with respect to density [the second virial coefficient of $S_{\tau, \tau'}(T,n_n,n_p)$].

Using the virial expansion of the density EoS (\ref{bnnp}) we can also perform the virial expansion of the incompressibility (\ref{incompr}) and, using Eq. (\ref{fi0}), for the excess quantity $\varphi_0(T,n)$. For symmetric matter ($Y_p=0.5)$ we obtain the virial expansion
\begin{equation}
\label{virphi}
 \varphi_0(T,n)=-\frac{n \Lambda^3}{2}(b_{nn}+b_{np}-b_{nn}^{(0)}) + {\cal O}(n^2)= \varphi^{(1,{\rm sym})}_0(T)n+ {\cal O}(n^2),
\end{equation}
where $b_{nn}^{(0)}=-2^{-5/2}$ is the contribution of the ideal Fermi gas. Using the second virial coefficients of Ref. \cite{HS}, values for $\varphi^{(1)}_0(T)$ are given in Tab. \ref{tab:1}.
We add the corresponding results for neutron matter ($Y_p=0$) 
 \begin{equation}
\label{virphin}
\varphi^{(1,{\rm neut})}_0(T)=-\Lambda^3\left(b_{nn}+\frac{1}{2^{5/2}}\right)\,.
 \end{equation}

\begin{table}[ht]
\begin{tabular}{|c|c|c|c|c|c|c|}
\hline
$T$ & $\varphi^{(1,{\rm sym})}_0(T)$ [fm$^3$]& $\varphi^{(1,{\rm neut})}_0(T)$  [fm$^3$]\\
\hline
1& -41785.9  & -1954.2 \\
2& -4888.2 & -713.2 \\
3& -1815.3 & -390.6 \\
4& -966.2 & -254.3 \\
5& -606.7 & -182.3 \\
6& -421.5 & -138.7 \\
7& -311.8 & -110.1 \\
8& -241.3 & -90.26 \\
9& -193.6 & -75.80 \\
10& -159.3 & -64.85\\
\hline
\end{tabular}
\caption{\label{tab:1}%
Excess virial term $\varphi^{(1)}_0(T)$, Eq. (\ref{virphi}), of the incompressibility (\ref{fi0}) of nuclear matter
for different values of $T$. Results for symmetric matter $\varphi^{(1,{\rm sym})}_0(T)$ and neutron matter $\varphi^{(1,{\rm neut})}_0(T)$ using the virial coefficients of \cite{HS}.}
\end{table}

The second virial coefficient is a benchmark for the low-density behavior of $n(T, \mu)$.
The EoS (\ref{BU}) is derived within the Green function approach
if the  self-energy is taken in the two-nucleon ladder (binary collision) approximation.
It can be extended to higher densities if the medium modifications are taken into account,
for instance in mean-field QPA \cite{SRS}.
In particular,  disappearance of
bound states at increasing density because of Pauli blocking is described, see \cite{R}.
Within the real-time Green function technique, the corresponding improvements are described in \ref{sec:quasi}.

\subsection{Derivation of the Beth-Uhlenbeck formula from the dynamic structure factor}

\begin{figure}[!th]
\includegraphics[width=16cm]{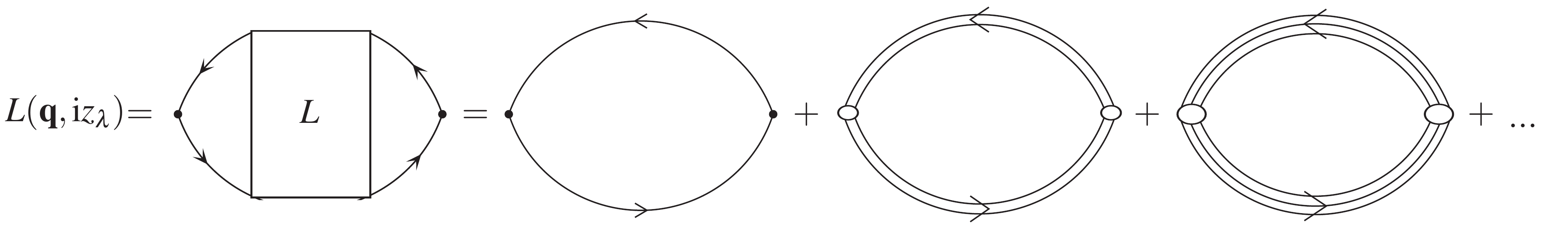}
\caption{Cluster decomposition of the polarization function. Cluster-particle propagation
is symbolized by the multiple particle propagators, see Eq. (\ref{G2op}),
and the vertex is introduced according to Eq. (\ref{Matrix}).}
\label{fig:Lcl}
\end{figure}

After demonstrating firstly the inclusion of cluster formation via the normalization condition (\ref{neos}), we investigate a second approach to the EoS starting from the dynamic structure factor calculating the isothermal compressibility (\ref{isot}).
A cluster decomposition of the van Hove function $L$ (\ref{Spectral}) can be performed which gives the possibility to include the
contribution of clusters (nuclei). This can be done as shown in \cite{RD,RMA93}.
To reproduce the Beth-Uhlenbeck formula which accurately describes the contribution of two-nucleon correlations in the low-density limit,
we search for the  diagrams  in the evaluation of the density-density correlation function which are necessary to reproduce this result. Higher order terms of interaction in the perturbation expansion, see Fig.~\ref{fig:1a}, will not produce bound state contributions.
To account for the formation of bound states, we have to add diagrams
where the single nucleon propagators in the one-loop diagram, Fig.~\ref{fig:1}, are replaced
by two-nucleon propagators as described by ladder sums \cite{RD,RMA93}. This leads us to
the cluster decomposition of the density-density Green function given tentatively in Fig. \ref{fig:Lcl}, in
analogy to the cluster decomposition of the self-energy, Fig. \ref{fig:SE}.
The loop diagram formed by free single-particle propagators (\ref{Lo}),
which describes the low-density limit,  is completed by loop diagrams
consisting of the two-particle propagator (\ref{G2op}), the three-particle propagator, etc.,
which are given by the corresponding ladder Green functions $G_A$. We find
\begin{equation}
\label{Lclu}
L({\bf q},\ii z_\lambda)=L_0({\bf q},\ii z_\lambda)+L^{(0)}_2({\bf q},\ii z_\lambda)+\dots
\end{equation}
The first-order contributions of the perturbation expansion $L_1({\bf q},z)$, see Fig. \ref{fig:1a},
are contained in $L^{(0)}_2({\bf q},z)$.
An accurate description of the cluster decomposition has to avoid unconnected diagrams and double counting.
A more detailed discussion is given below.

\begin{figure}[!th]
\includegraphics[width=8cm]{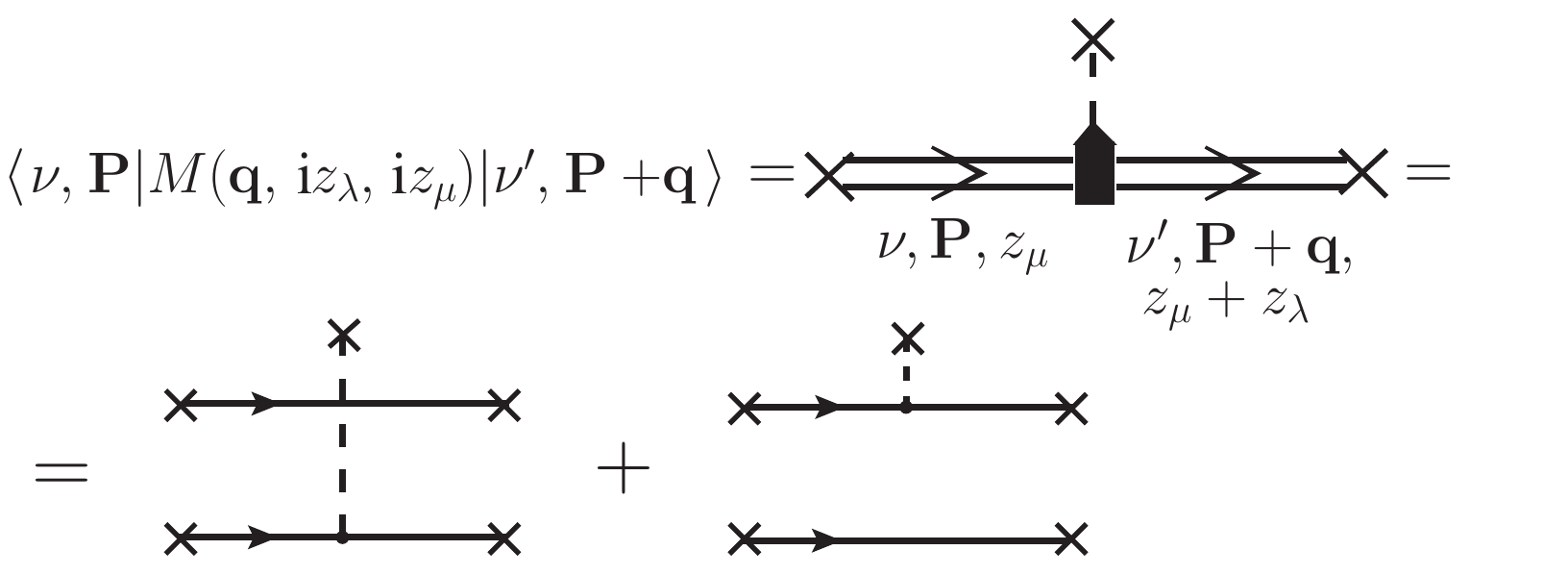}
\caption{Vertex matrix element for the two-particle contribution $L^{(0)}_2({\bf q},\ii z_\lambda)$
to the polarization function.}
\label{fig:3}
\end{figure}

We consider here only the contribution of two-particle correlations $L^{(0)}_2({\bf q},\ii z_\lambda)$.
The two-particle propagator in eigen-representation reads
\begin{equation}
\label{clusterprop}
\langle \nu,{\bf P}|G_2(z)|\nu',{\bf P}' \rangle = \frac{1}{z-E^0_{\nu,P}} \delta_{\nu\nu'} \delta_{{\bf P},{\bf P}'}.
\end{equation}
The two-particle vertex which is related to the density fluctuation describes, for instance,
the coupling to the interaction propagator. The coupling of the cluster to the interaction is given by the matrix element given in Fig. \ref{fig:3}. The crosses denote amputation, i.e. multiplication with $[G_2^{(0)}]^{-1}$.
Using the two-nucleon eigen-states $|\nu,{\bf P}\rangle$ with intrinsic quantum number $\nu$ and c.m. momentum ${\bf P}$, we define this matrix element according to
\begin{equation}
\label{Matrix}
M_{\nu \nu'}({\bf q})=\langle \nu,{\bf P}|M({\bf q}, \ii z_\lambda, \ii z_\mu) |\nu',{\bf P}+{\bf q} \rangle = \sum_{{\bf p}_1,{\bf p}_2} \psi^*_{\nu,{\bf P}}(p_1,p_2)
[\psi_{\nu',{\bf P}+{\bf q}}({\bf p}_1+{\bf q},{\bf p}_2)+\psi_{\nu', {\bf P}+{\bf q}}({\bf p}_1,{\bf p}_2+{\bf q})].
\end{equation}

\begin{figure}[!th]
\includegraphics[width=8cm]{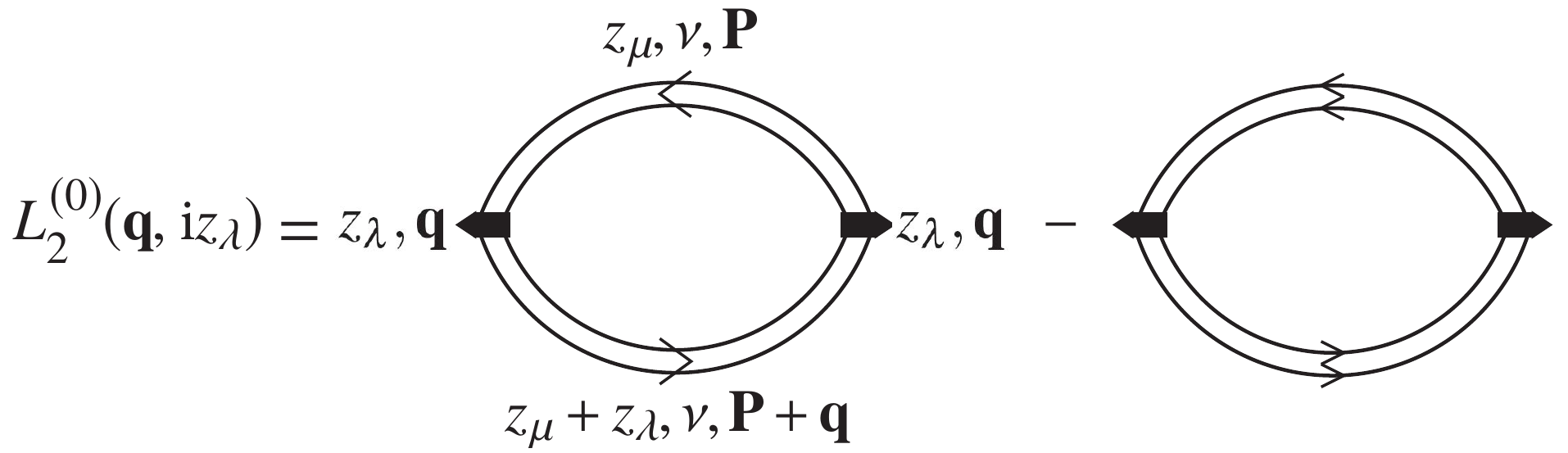}
\caption{Two-particle contribution $L^{(0)}_2({\bf q},\ii z_\lambda)$ to the polarization function. The cluster propagator is given by Eq. (\ref{clusterprop}). The free two-particle contribution
has to be subtracted because it is a disconnected diagram.}
\label{fig:2}
\end{figure}

For the two-particle contribution to the van Hove function we obtain (see Fig. \ref{fig:2})
\begin{equation}
L^{(0)}_2({\bf q}, z)=\sum_{\nu\nu',{\bf P}}\frac{f_{\rm B}(\epsilon^0_{\nu,{\bf P}})-f_{\rm B}(\epsilon^0_{\nu', {\bf P}+{\bf q}})}{z+\epsilon^0_{\nu,{\bf P}}-\epsilon^0_{\nu', {\bf P}+{\bf q}}}|M_{\nu\nu'}({\bf q})|^2 -\sum_{{\bf p}_1,{\bf p}_2}
\frac{f_{\rm B}(\epsilon^0_{p_1}+\epsilon^0_{p_2})-f_{\rm B}(\epsilon^0_{{\bf p}_1+{\bf q}}+\epsilon^0_{p_2})}{z+\epsilon^0_{p_1}-\epsilon^0_{{\bf p}_1-{\bf q}}}\,.
\end{equation}

After inserting this contribution in the expression for the dynamic structure factor as discussed for the single nucleon contribution,
we transform $[f_{\rm B}(\epsilon^0_{\nu,P})-f_{\rm B}(\epsilon^0_{\nu',{\bf P}+{\bf q}})]f_{\rm B}(\epsilon^0_{\nu,P}-\epsilon^0_{\nu',{\bf P}+{\bf q}})=f_{\rm B}(\epsilon^0_{\nu,P}) [1+f_{\rm B}(\epsilon^0_{\nu',{\bf P}+{\bf q}})]$.
In the  limit ${\bf q} \to 0$ the expression $f_{\rm B}(\epsilon^0_{\nu,P}) [1+f_{\rm B}(\epsilon^0_{\nu',P})]$ follows.
The matrix elements (\ref{Matrix}) are simplified in the limit ${\bf q} \to 0$. Using the completeness relation $ \sum_{{\bf p}_1,{\bf p}_2} |{\bf p}_1,{\bf p}_2 \rangle \langle {\bf p}_1,{\bf p}_2 |= {\rm I}$  the summation over ${\bf p}_1,{\bf p}_2$
gives
\begin{equation}
\lim_{{\bf q} \to 0}M_{\nu\nu'}({\bf q})=2 \langle \nu,{\bf P}|\nu',{\bf P}\rangle=2 \delta_{\nu,\nu'}.
\end{equation}

The two-nucleon contribution to the dynamic structure factor (\ref{SL}) and the   isothermal compressibility (\ref{isot}) is calculated
using
\begin{equation}
\label{iso2}
\kappa^{(2)}_{\rm iso}(T,\mu)
=\frac{1}{n^2T}\left\{ \sum_{\nu,{\bf P}} f_{\rm B}(\epsilon^0_{\nu,P}) [1+f_{\rm B}(\epsilon^0_{\nu,P})] -\sum_{{\bf p}_1,{\bf p}_2} f_{\rm B}(\epsilon^0_{p_1}+\epsilon^0_{p_2}) [1+ f_{\rm B}(\epsilon^0_{p_1}+\epsilon^0_{p_2})]\right\}\,.
\end{equation}
Quite similar as in the derivation of the Beth-Uhlenbeck formula in Sec. \ref{sec:BU}, the second contribution
owing to the free nucleon states compensates the divergent part of the scattering states in the first contribution of Eq. (\ref{iso2}),
caused by the disconnected diagrams arising from the zeroth order of the ladder sum for the two-particle propagator.
Selecting only the bound state part of $\kappa^{(2)}_{\rm iso}(T,\mu)$, a mass action law is obtained as known from the NSE.
As demonstrated in Sec. \ref{sec:BU}, the summation over $\nu$ can be replaced by an integral over $E_{\rm rel}$ where the density of
states $D^{\rm BU}_{\alpha, {\bf P}}(E_{\rm rel})$ (\ref{DBU}) appears, and the summation over the (spin, isospin) channels $\alpha$.

The following expression for the isothermal compressibility is observed
\begin{eqnarray}
\label{kapBU}
&&\kappa^{(\rm BU)}_{\rm iso}(T,\mu_n,\mu_p)=\frac{1}{V n^2T}
\left\{ \sum_{\bf p} f(\epsilon_p^0) (1-f(\epsilon_p^0))
+ \sum_{\alpha, {\bf P}} \int_{-\infty}^\infty \frac{d E}{\pi} f_{\rm B}\left( \epsilon^0_{\alpha,P} \right) \left[1+f_{\rm B}\left(\epsilon^0_{\alpha,P} \right)\right]
D^{\rm BU}_{\alpha, {\bf P}}(E)\right\}\,.
\end{eqnarray} 
It is easily shown that  this expression is consistent with the Beth-Uhlenbeck formula (\ref{BU}) given above.

\subsection{The nuclear statistical equilibrium model}
\label{sec:NSE}

Starting from the dynamic structure factor and the isothermal compressibility, we presented the way
how to implement the formation of clusters. The cluster decomposition of the polarization function $L({\bf q}, z) \approx L_0({\bf q}, z)+L^{(0)}_2({\bf q}, z)+\dots+L^{(0)}_A({\bf q}, z)$
gives the possibility to include also larger clusters, i.e. in addition to the deuteron also the other light elements $^3$H, $^3$He, and $^4$He as well as larger clusters, characterized by the mass number $A$ and the intrinsic quantum state $\nu$
(including spin and isospin as well as excitation level).
Note that this cluster decomposition gives the correct virial limit if in addition to the bound states also scattering states are taken into account. For instance, $L^{(0)}_2({\bf q}, z)$
produces also the mean-field quasiparticle shifts shown in Fig. \ref{fig:1a}.

We have to consider the contribution $L^{(0)}_A({\bf q}, z)$ containing the $A$-particle propagator. This leads to the corresponding $A$-nucleon in-medium wave equation, see \cite{R}. We will consider here only the low-density limit where the solution of the
$A$-nucleon wave equation gives as bound states the nuclei with mass number $A$.
We also restrict us to only the bound state contribution neglecting contributions of the continuum.

The calculation of the compressibility is along to the former case, Eqs. (\ref{eq:22}), (\ref{iso2})  and gives
\begin{eqnarray}
\kappa^{(\rm NSE)}_{\rm iso}(T,\mu_n,\mu_p)&=&\frac{1}{V n^2T}
\left\{ \sum_{A=1,3,\dots}\sum_{\nu}^{\rm bound}g_{A,\nu} \sum_{\bf P} f(\epsilon^0_{A,\nu,{ P}}) [1- f(\epsilon^0_{A,\nu,{ P}})]
\right. \nonumber \\ && \left. + \sum_{A=2,4,\dots}\sum_{\nu}^{\rm bound} g_{A,\nu} \sum_{\bf P} f_{\rm B}(\epsilon^0_{A,\nu,{ P}}) [1+ f_{\rm B}(\epsilon^0_{A,\nu,{ P}})]\right\}\,.
\end{eqnarray}
with the Fermi function for odd numbers $A$ and the Bose function for even $A$.
$g_{A\nu}$ denotes the degeneracy of the cluster with mass number $A$ and the intrinsic quantum state $\nu$.
The binding energies $ E^0_{A,\nu,{ P}} $
of that clusters determine $\epsilon^0_{A,\nu,{ P}} =E^0_{A,\nu,{ P}}-Z \mu_p-(A-Z) \mu_n$, $Z$ denotes the proton number.
It is easily shown that this result corresponds to the standard result \cite{RMS}
\begin{eqnarray}
\label{kapNSE}
n^{(\rm NSE)}(T,\mu_n,\mu_p)&=&\frac{1}{V }
\sum_{A=1,3,\dots}\sum_{\nu}^{\rm bound}g_{A,\nu} \sum_{\bf P} f(\epsilon^0_{A,\nu,{ P}})
+\frac{1}{V }\sum_{A=2,4,\dots}\sum_{\nu}^{\rm bound} g_{A,\nu} \sum_{\bf P} f_{\rm B}(\epsilon^0_{A,\nu,{ P}}).
\end{eqnarray} 
Thus we are convinced that clusters are correctly implemented in the alternative approach which is based on
the evaluation of the dynamic structure factor. Not the improvement of the single nucleon quasiparticle approach,
but the cluster expansion of the density-density Green function leads to the consistent treatment of bound nuclei.

The inclusion of scattering states becomes complex for $A >2$, see \cite{R}.
A cluster-virial expansion which treats the cluster-scattering states has been discussed in Ref. \cite{clustervirial}.

\section{Compressibility including cluster formation}
\label{comp}

Considering the isothermal compressibility as the key quantity of the present work,
we find the expression $\kappa^{(\rm BU)}_{\rm iso}(T,\mu_n,\mu_p)$ (\ref{kapBU})
in the low-density limit which tells us that the cluster contributions enter additively
the low-density limit, see also Eq. (\ref{kapNSE}).
A challenge is the extension of the cluster decomposition of the density-density Green function to higher densities where
self-energy and Pauli blocking must be included. We will not investigate this problem here but give only some brief comments.

For the single-nucleon contribution (first term in Eq. (\ref{kapBU})), the quasiparticle picture, Sec. \ref{sec:quasi}, can be introduced which allows to describe also
nuclear systems
near the saturation density.
This quasiparticle concept has been proven to be very efficient to describe matter at high densities.
It has been demonstrated in Sec. \ref{sec:mean} that the quasiparticle concept can also be introduced in the dynamic structure factor approach derived from the density-density Green function. However, in comparison with the
density EoS approach (\ref{neos}), more effort is needed within the density-density Green function approach, see Sec. \ref{sec:mean}, and in addition to the self-energy, also vertex terms have to be considered.
However, the latter approach opens the access to further quantities such as dynamic properties and transport processes. As pointed out in this work, similar concepts known from the density-EoS approach (\ref{neos})
can also applied to the cluster decomposition of $L$ (\ref{Lclu}) so that both approaches are equivalent.

Of interest is what we expect for the compressibility if cluster formation is taken into account.
We give some results valid for dilute warm nuclear matter according to Ref. \cite{R}.
The solution of the effective $A$-nucleon in medium wave equation leads to medium-dependent shifts of the bound state energies.
This can be interpreted as medium-dependent quasiparticle energies of the bound states.

Starting from the normalization condition which gives the density as function of $T, \mu_\tau$,
a generalized Beth-Uhlenbeck formula accounting for in-medium corrections was found \cite{SRS}
which includes quasiparticle-like  bound states
 as well as in-medium scattering states,
\begin{eqnarray}
\label{kapgBU}
&&\kappa^{(\rm BU)}_{\rm iso}(T,\mu_n,\mu_p)=\frac{1}{V n^2T}
\left\{ \sum_{\bf p} f(\epsilon_p) (1-f(\epsilon_p))
+ \sum_{\alpha, {\bf P}} \int_{-\infty}^\infty \frac{d E}{\pi} f_{\rm B}\left( \epsilon_{\alpha,P} \right) \left[1+f_{\rm B}\left(\epsilon_{\alpha,P} \right)\right]
D^{(\rm gBU)}_{\alpha, {\bf P}}(E)\right\}\,
\end{eqnarray}
with (after partial integration and using the Levinson theorem, see \cite{R})
\begin{equation}
\label{DgBU}
 D^{\rm gBU}_{\alpha , {\bf P}}(E_{\rm rel})=g_\alpha \frac{1}{T}\left(\sum^{\rm bound}_\nu \pi [\delta(E_{\rm rel}-E_{\alpha \nu { P}})-\delta(E_{\rm rel})]
 +[ \delta_{\alpha, { P}}(E_{\rm rel})-\frac{1}{2}\sin(2\delta_{\alpha ,{ P}}(E_{\rm rel}))] \right)\,.
\end{equation}
In contrast to the ordinary Beth-Uhlenbeck formula (\ref{kapBU}),
the free single particle energies $E^0_p$ are replaced by the quasiparticle energies $E_p(T,n,Y_p)$
which are depending on temperature and densities.
However, caution is needed to avoid double counting.
If low-order contributions of the ladder sum have been used already
to define the quasiparticles such as the Hartree-Fock shifts,
they have to be eliminated from the two-particle contributions.
This is the reason for the appearance of the sin-term at the end of Eq. (\ref{DgBU})
which is obtained from a consistent treatment of the optical theorem  \cite{SRS}.
In addition, the generalized Beth-Uhlenbeck formula contains also medium-modified
binding energies $E_{\alpha \nu { P}}(T,n,Y_p)$ and scattering phase shifts $\delta_{\alpha ,{ P}}(E_{\rm rel};T,n,Y_p) $ depending on $T,n_\tau$ which are obtained from an in-medium Schr{\"o}dinger equation.
An important fact is the disappearance of bound states at increasing density because of Pauli blocking.

\begin{figure}[!th]
\includegraphics[width=10cm]{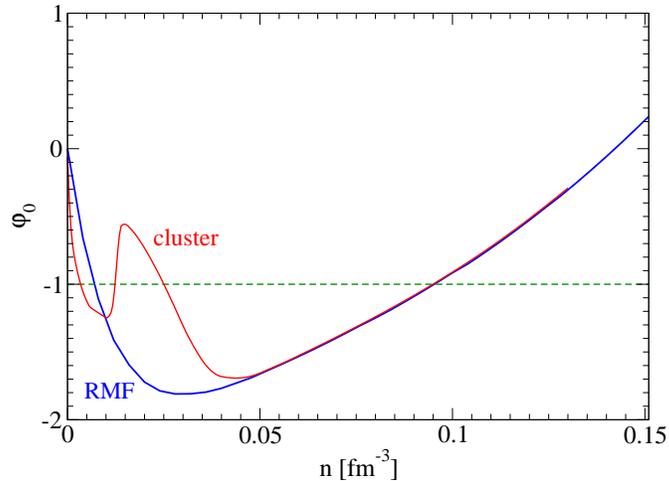}
\caption{Excess contribution $\varphi_0(T, n)$ to the incompressibility, Eq. (\ref{fi0}),  at $T=5$ MeV for symmetric matter ($Y_p=0.5$). The DD2-RMF approximation, see App. \ref{DD2}, is compared with the result accounting for light cluster formation \cite{R}.}
\label{fig:9}
\end{figure}

Using the density EoS of dilute warm nuclear matter obtained in \cite{R},
results for the excess contribution to the incompressibility  $\varphi_0(T,n)$  (\ref{fi0}) are shown in Fig. \ref{fig:9}  for $T=5$ MeV as function of the density. The region of spinodal instability is given
by the condition $\varphi_0(T,n) < -1$. Within the RMF approximation, a smooth behavior is obtained, the region of instability is shown. For zero temperature, in this RMF approximation the result for $f_0$ shown in Fig. \ref{fig:0}
is obtained. To investigate the influence of light cluster ($d, t, h, \alpha$) formation, the excess quantity
$\varphi_0(T,n)$ is also shown as obtained from the quantum statistical approach, extending the generalized
Beth-Uhlenbeck formula (\ref{kapgBU}) by including all cluster with $A\le 4$, see Ref. \cite{R}.

Strong deviations are  found in the low-density region owing to the formation and dissolution of clusters. This is already seen from the virial expansion (\ref{virphi}) of the excess quantity $\varphi_0(T,n)$.
The evaluation of the term linear in the density for symmetric matter at $T=5$ MeV within the DD2-RMF approximation
(see Appendix \ref{DD2}) gives $\varphi_0^{\rm RMF}(T=5\, {\rm MeV},n)=-165.5\, n\, {\rm fm}^3+{\cal O}(n^2)$.
This value is only slightly different from the zero-temperature result $\varphi_0^{\rm RMF}(T=0,n)=-202.4\, n\, {\rm fm}^3+{\cal O}(n^2)$. An accurate result for symmetric matter at $T=5$ MeV,
\begin{equation}
 \varphi_0(T=5\,{\rm MeV},n)=-606.73\, n\, {\rm fm}^3+{\cal O}(n^2)
\end{equation}
is obtained using the data for the second virial coefficients \cite{HS}, see also Tab. \ref{tab:1}. As shown in Fig. \ref{fig:9}, the RMF QPA fails to reproduce this benchmark.

In particular, two regions of spinodal instability are obtained, and in between a region of metastability occurs.
This result is also directly seen from the   EoSs shown in  \cite{R}. The influence of cluster formation is significant for densities below 0.05 fm$^{-3}$. The mass fraction of $\alpha$ particles is large near the baryon density 0.01 fm$^{-3}$, see \cite{R} for the composition at $T=5$ MeV. At higher densities the bound states disappear because of Pauli blocking, and the RMF QPA is applicable.

Coming back to the second approach to the thermodynamics of nuclear matter,
the investigation of the dynamic structure factor is expected to give results
for the compressibility  identical with the density EoS approach (\ref{neos}) as shown in the low-density limit, but more effort is needed to treat the corrections at higher density. However, this alternative approach is able to give interesting properties with respect to the dynamic behavior of nuclear systems
not discussed in the present work.

\section{Discussion and conclusions}
\label{Conclusions}

Our aim is to find the relation between the well-established Landau Fermi-liquid approach to nuclear systems and other
approaches such as the QS approach based on the normalization condition. In particular, we are interested in the problem how
the formation of light clusters such as $d,t,h,\alpha$ can be described. Previous work such as \cite{KV16} pointed out this problem.
There, it was not clear how the formation of bound states can be included in an approach where nucleons are described as quasiparticles.

To solve this problem we propose to consider a fundamental quantity, the density-density correlation function which is related to the
dynamic structure factor. The compressibility is included as limiting case and can be used to derive the nuclear matter EoS.
In addition, several other properties are related to the dynamic structure factor, equilibrium as well as non-equilibrium. Therefore, the density-density correlation function is an
 important quantity by itself, and the systematic quantum-statistical treatment of cluster formation is an
 actual problem. We found a solution of this problem performing a cluster decomposition of the density-density correlation function, i.e. using the concept of the chemical picture. Ladder  sums describing few-body correlations are
considered as new element of a diagram representation of the thermodynamic Green functions. A cluster decomposition of
the relevant quantities, such as the self-energy or the polarization function, allows for the description of bound-state formation.
We show that this alternative approach leads to the same results known from the QS approach using the density EoS (\ref{neos}).
In particular, the exact results for the second virial coefficient, the Beth-Uhlenbeck  formula, are obtained in both approaches.

Considering the isothermal compressibility as the key quantity of the present work,
we find the expression $\kappa^{(\rm BU)}_{\rm iso}(T,\mu_n,\mu_p)$ (\ref{kapBU})
in the low-density limit which tells us that the cluster contributions enter additively
the low-density limit, see also Eq. (\ref{kapNSE}). The excess contribution to the incompressibility  $\varphi_0(T,n)$  (\ref{fi0}) is an
 important quantity describing the interaction effects as well as the cluster formation in nuclear matter.

The relation to the Landau Fermi-liquid approach is found as follows: In the high-density (near the saturation density),
zero temperature case the nuclear system is degenerate, correlations are taken into account by the quasiparticle picture,
and bound state formation is suppressed because of Pauli blocking.
The dynamic structure factor and the isothermal compressibility are obtained from the  QPA described in
Sec. \ref{sec:quasi}. Instead of the knowledge of the full interaction, only special Landau-Migdal parameters are needed.
We gave the parameter values comparing with well accepted parametrizations
of nuclear matter properties within the RMF, in particular DD2.

However, to include bound state formation what is essential in the low-density region,
we have to go beyond the Landau Fermi-liquid approach. Further diagrams for the dynamic
structure factor must be considered which appear in a cluster decomposition of the polarization function. They can be added to the
quasiparticle contribution to the isothermal compressibility. However, double counting has to be avoided.

Similar expressions may also be derived from the density-density Green function
approach. The medium-modified solutions of the two-nucleon problem have to be implemented in the cluster decomposition of the polarization function.
The relation of the single-quasiparticle contribution to the Fermi-liquid approach opens possibilities to treat non-equilibrium, inhomogeneous
processes in nuclear systems so that it is of relevance to have a consistent
description of equilibrium  properties. However, this is sufficient only for parameter values
where cluster formation can be neglected. We have shown that additional contributions to the polarization function are obtained considering the two-particle clusters.
In particular, we have shown how the second virial coefficient appears.

It is rather difficult to include density effects for arbitrary cluster size $A$.
We expect similar results as the generalized Beth-Uhlenbeck formula.
It was not the aim of this work
to reproduce all sophisticated results obtained in the QS approach \cite{R} until now,
starting from the normalization condition (\ref{neos}).
We only demonstrated how to proceed with the density-density Green function approach to include
bound state formation, and standard results were reproduced like the second virial coefficient.
This is of interest because the polarization function is related to many physical properties,
including excitations and transport properties. In particular, the dynamic structure factor
is obtained, and it has been shown how the influence of clustering can be considered 
using the cluster decomposition of the polarization function. \\

{\bf Acknowledgements.}  D.N.V. and D.B. were  supported  by  the
Russian  Science  Foundation,  Grant  No.    17-12-01427. D.N.V.  was also supported by the Ministry of Education and Science of the Russian Federation within the state assignment, project No 3.6062.2017/BY.\\

\appendix

\section{Relations between two-point functions in non-equilibrium diagram technique}
\label{A1}

For any two-point function
 \begin{eqnarray} \label{notation}
 F_{i}^{\,\,j}(x,y)=\sigma_{ik}F^{kj}(x,y),\quad F^{i}_{\,\,j}(x,y)=F^{ik}(x,y)\sigma_{kj},
 \quad F_{ij}=\sigma_{ik}\sigma_{jl}F^{kl},\quad \sigma_{i}^{\,\,k}=\delta_i^{\,\,k}\,,
 \end{eqnarray}
 $\{i,k\}$ are $-$ or $+$.

 In equilibrium all non-equilibrium boson  self energies and fermion Green functions can be expressed via the retarded ones, c.f. \cite{Ivanov:1999tj}, e.g.,
 \begin{eqnarray}\label{eqrel}
 \ii \Pi^{-+}(q_0,{\bf{q}}) =f_{\rm B}(q_0)\Gamma_{\rm B} (q_0,{\bf{q}})\,,\quad
 \ii \Pi^{+-}(q_0,{\bf{q}}) =(1+f_{\rm B}(q_0))\Gamma_{\rm B} (q_0,{\bf{q}})\,,
 \end{eqnarray}
 \begin{eqnarray}\label{eqrel1}
 -\ii G^{-+}(p_0,{\bf{p}}) =f(p_0-\mu)A (p_0,{\bf{p}})\,,
 \quad \ii G^{-+}(p_0,{\bf{p}}) =(1-f(p_0-\mu))A (p_0,{\bf{p}})\,,
 \end{eqnarray}
 with
 \begin{eqnarray}\label{Gfb}
 &\Gamma_{\rm B}=-2 {\rm Im}\,\Pi^R\,,
  &A=-2 {\rm Im}\,G^R=\frac{\Gamma}{[p_0-E_p^0-\mbox{Re}\Sigma^R(p_0, {\bf p})]^2+\Gamma^2/4},\quad \Gamma =-2 {\rm Im}\,\Sigma^R\,,
  \end{eqnarray}
   and thermal distributions
 \begin{eqnarray}\label{thdistr}
 f_{\rm B}(q_0)=\frac{1}{e^{q_0/T}-1}\,,\quad f(p_0-\mu)=\frac{1}{e^{(p_0-\mu)/T}+1}\,,
\end{eqnarray}
$\mu (T,n)$ is the fermion chemical potential.
 We present also a helpful relation between thermal distributions, see (\ref{f1-f}),
 \begin{eqnarray}\label{threl}
 f(p_0-\mu +q_0) (1-f(p_0-\mu) )=[f(p_0-\mu)-f(p_0-\mu +q_0)] f_{\rm B} (q_0)\,.
\end{eqnarray}

\section{The single-particle spectral function in the real-time Green function approach}
\label{sec:spectral}

\subsection{One-loop result with full  Green functions}

After demonstrating the Matsubara Green function approach, let us consider the real-time Green function method introduced in Sec. \ref{sec:nonequGF}.
In some cases it is helpful to expand $\Pi^{-+}$ in a series of $(G^{-+}G^{+-})$ loops
with full Green functions, cf. \cite{Knoll:1995nz}.
Consider the first ($N=1$) diagram of $\Pi^{-+}$ with
the full fermion propagators
\begin{equation}\label{one-loop}
\left[\unitlength6mm\begin{picture}(3.5,1)
   \put(0,0.15){\oneloopvertex}\end{picture}\right]^{00}
   =-\ii\Pi_{-+}^{N=1,00}(q;X)=S^{N=1}(q;X)\,,
\end{equation}
\begin{eqnarray}
S^{N=1}({\bf{q}}\to 0;X)=\int\frac{\di q_0}{2\pi}[-i\Pi_{-+}^{N=1,00} (q;X)]\,.
\end{eqnarray}

In the one $- +$ loop ($N=1)$ approximation with fully dressed Green functions from (\ref{eqrel1}) using (\ref{threl}) and also using that $\Gamma \neq 0$  for $|{\bf{q}}|>q_0$ we find
\begin{eqnarray}\label{barestT1}
S_0^{N=1}({\bf{q}}\to 0) &=&T\int\frac{\di q_0}{2\pi q_0}g\frac{\di^4 p}{(2\pi)^4}A(p_0+q_0, {\bf{p}}) A(p_0 , {\bf{p}})(f(p_0-\mu)-f(p_0 +q_0 -\mu))\nonumber\\
&=&\mbox{Re}\Pi^R_0 (0,{\bf{q}}\to 0)\,,
\end{eqnarray}

\subsection{Quasiparticle and improved quasiparticle approximations}
\label{sec:quasi}
The quasiparticle approximation (QPA) is a commonly used concept originally derived for Fermi liquids at low temperatures, see \cite{LL,Mig,Mig1}, where it
constitutes a consistent approximation scheme.  For the validity of the QPA one normally assumes that
the fermion width $\Gamma\ll\bar{\epsilon}$, where $\bar{\epsilon}\sim T$ in equilibrium matter, cf. (\ref{Gfb}). Besides, one needs to demand that $q_0\gg \Gamma$ in the QPA \cite{Knoll:1995nz}. For $T\ll \epsilon_{\rm F}$, $\Gamma\sim \pi^2 T^2/\epsilon_{\rm F}$ and the QPA is applicable for  $q_0\gg \Gamma$.

We start from the single-nucleon propagator and its spectral function
$A (p_0,{\bf p})$. Within the QPA, the spectral function is replaced by a $\delta$ function at the quasiparticle energy. The use of QPA gives
considerable computational advantages since the Wigner densities ("-- +" and
"+ --" Green functions) become energy $\delta$--functions. Then, the particle
occupations  depend only on the momentum  rather than on
the energy variable. Formally the energy integrals are
eliminated in diagrammatic terms  cutting the corresponding "--
+" and "+ --" lines \cite{Voskresensky:1987hm}. The quasiparticle picture allows a transparent interpretation of closed
diagrams. Thus, in the QPA
\begin{eqnarray}
A^{\rm QPA}(p_0,{\bf p})=2\pi \delta (p_0 - E_{p}^0 -{\rm Re}\Sigma^R (p_0,{\bf p}))=\left(1-\frac{\partial {\rm Re}\Sigma^R}{\partial p_0}\right)_{E_p,{\bf{p}}}^{-1}\delta (p_0 -E_p) \,.
\end{eqnarray}
The quasiparticle energy $E_p$ is the root
 of the condition
 \begin{eqnarray}\label{spectr}
 p_0 - E_{p}^{\,0} -{\rm Re}\Sigma^R (p_0,{\bf{p}})=0\,,
 \end{eqnarray}
 where $\Sigma(p_0,{\bf p})$  is the fermion  self-energy. The residue $\left(1-\partial {\rm Re}\Sigma^R/\partial p_0\right)_{E_p,{\bf p}}^{-1}$ is positive-definite provided the particle-antiparticle separation is appropriately performed. In the relativistic case we have $E_{p}^{\,0} =\left({\bf{p}}^{\,2} +m^{\,2}\right)^{1/2}$.
 In the presence of the vector field $U=(U_0, {\bf U})$, like for the nucleon interacting with the $\omega$ meson, we should still perform the shift $\mu\to\mu-U_0$.  Note that $A^{\rm QPA}$ does not satisfy the exact sum-rule but
 \begin{eqnarray}\int_{-\infty}^{\infty}A^{\rm QPA} dp_0/(2\pi) =\left(1-\frac{\partial {\rm Re}\Sigma^R}{\partial p_0}\right)_{E_p,{\bf{p}}}\,
\end{eqnarray}
and the necessity of the renormalization arises, if one works within the QPA follows.

Now consider an ``improved" quasiparticle approximation (IQPA) \cite{SRS}. For $|\partial {\rm Re}\Sigma^R/\partial p_0|\ll 1$, the expansion up to the linear term in $\Gamma$ gives
\begin{eqnarray}\label{IQPA1}
A^{\rm IQPA}(p_0,{\bf p})=-2 {\rm Im} \left[p_0 - E_{p}^{\,0} -\Sigma^R (p_0,{\bf{p}})\right]^{-1}
\simeq \left(1+\frac{\partial {\rm Re}\Sigma^R}{\partial p_0}\right)_{E_p,{\bf{p}}}\delta (p_0 -E_p)+\Gamma {\cal{P}} \frac{1}{(p_0-E_p)}+O(\Gamma^2) \,.
\end{eqnarray}
The symbol ${\cal{P}}$ means the principal value. Inserting (\ref{IQPA1}) in
(\ref{barestT1}) and (\ref{EQdistr}) and using the Kramers-Kronig relation
\begin{eqnarray}\label{KramersKronig}
\left(
\frac{\partial {\rm Re}\Sigma^R}{\partial p_0}\right)_{E_p,{\bf{p}}}=-{\cal{P}}
\int_{-\infty}^{\infty}\frac{\Gamma (\omega')\di \omega'}{2\pi(E_p-\omega')^2},
\end{eqnarray}
 and we see that $A^{\rm IQPA}(p_0,{\bf p})$ obeys the exact sum-rule, namely
\begin{eqnarray}
\int_{-\infty}^{\infty} A^{\rm IQPA}(p_0,{\bf p})dp_0/(2\pi) =1\,.
\end{eqnarray}
Thus within the IQPA we arrive at the simple relations
\begin{eqnarray}\label{IQPAbarest1}
S_0^{N=1}({\bf{q}}\to 0) =g\int\frac{\di^3 p}{(2\pi)^3} f(E_p-\mu) [1- f(E_p-\mu)]
+{\cal O}(\Gamma^2)\,,
\end{eqnarray}
 \begin{eqnarray}\label{IQPAdistr}
 n=g\int\frac{\di^3 p}{(2\pi)^3}  f(E_p-\mu)+{\cal O}(\Gamma^2)\,.
\end{eqnarray}
Note that within the self-consistent Hartree approximation,
$\Gamma\to 0$ and the IQPA works perfectly.

In non-relativistic approximation $|\omega=p_0-m|\ll m, |\vec{p}|\ll m$,   we  expand ${\rm Re}\Sigma^R (\omega,{\bf{p}})$ with respect to $\omega$ near the root of the dispersion law at ${\bf p}^2=0$ and
 ${\bf p}^2$ keeping only the constant and  ${\bf p}^2$ terms. Introducing the non-relativistic  Landau effective fermion mass
 \begin{eqnarray}
m^* =\frac{m\left(1-\partial \mbox{Re}\Sigma^R(\omega,{\bf{p}})/\partial \omega\right)_{\Delta,{\bf{0}}}}{\left(1+2m\partial \mbox{Re}\Sigma^R(\omega,{\bf{p}})/\partial {\bf p}^2\right)_{\Delta,{\bf{0}}}}
\end{eqnarray}
as solution of Eq. (\ref{spectr}) we have $p_0=E_{p}=\Delta+{\bf p}^{\,2}/(2m^{*})$ where
 \begin{eqnarray}
\Delta =\frac{\left(\mbox{Re}\Sigma^R(\omega,{\bf{p}})\right)_{\Delta,{\bf{0}}}}{\left(1-\partial \mbox{Re}\Sigma^R(\omega,{\bf{p}})/\partial \omega\right)_{\Delta,{\bf{0}}}}\,.
\end{eqnarray}
Assuming $|\mbox{Re}\Sigma^R|$ being small  we  get $E_p^{\rm pert}\simeq {\bf p}^2/2m +\mbox{Re}\Sigma^R(0,0).$

\subsection{Other approximations}
 The spectral function $A(p_0,{\bf p})$ is in general a complicated function of the chemical potentials $\mu_\tau$ due to dependence of $\Sigma^R (\mu_\tau)$. When the $\mu$ dependence in $A(p_0,{\bf p})$ can be neglected
taking derivative $\frac{\partial}{\partial n}$ from both sides of Eq. (\ref{EQdistr})  we  obtain
\begin{eqnarray}\label{dmun}
&&S({\bf q}\to 0)=
\frac{ h}{T+ h(\partial V_0/\partial n_B)}\,,\quad
\nonumber \\ && h=g\int\frac{\di^4 p}{(2\pi)^4}A(p_0,{\bf p}) f(p_0-\mu +V_0)(1-f(p_0-\mu +V_0))\,.
\end{eqnarray}

In perturbation theory, for $E_p\simeq E_p^0$ we get
\begin{eqnarray}\label{hpert}
h_{\rm pert}=g\int\frac{\di^3 p}{(2\pi)^3} f(E_p^0-\mu +V_0)(1-f(E_p^0-\mu +V_0))\,,
\end{eqnarray}
 cf.
 Eq. (\ref{IQPAbarest1}). With the help of (\ref{threl}) we rewrite
 \begin{eqnarray}\label{hpertR}
h_{\rm pert}=\lim_{q_0\to 0}g\int\frac{\di^3 p}{(2\pi)^3}
[f(E_p^0-\mu+V_0)-f(E_p^0-\mu +V_0 +q_0)] f_{\rm B} (q_0)=T\mbox{Re}\Pi^R_0 (0,{\bf q}\to 0)\,,
\end{eqnarray}
since the standard loop expression yields
 \begin{eqnarray}
\label{PiR0}
 \Pi^R_0 (q_0,{\bf q})=\int \frac{\di^3 p}{(2\pi)^3}\frac{f(E_p^0-\mu+V_0)-f(E_{{\bf p}+{\bf q}}^0-\mu +V_0 +q_0)}{q_0 +E_p^0 -E_{{\bf p}+{\bf q}}^0+\ii 0}\,.
 \end{eqnarray}

The interaction via the density dependent potential $V_0$ results in a resummation of the loops.
However, no phase transition or cluster formation is obtained from this lowest order  approximation with respect to the interaction.

\section{Parametrization of the DD2-RMF QPA}
\label{DD2}


 We use units MeV, fm, $m_N=939.17$ MeV, $p_{{\rm F},\tau} =(3 \pi^2 n_\tau[{\rm fm}^{-3}])^{1/3}\, 197.3$ MeV.
The thermodynamics in the RMF approach is given by the relations (\ref{RMF1}), (\ref{RMF2}).
Both the scalar  ($S_\tau$) and vector part ($V_\tau$) of the self-energy are obtained from an empirical meson-hadron Lagrangian with effective (density dependent)
coupling constants.

For direct use, a parametrization for the DD
model \cite{Typel1999} was presented in Ref.\
\cite{Typel,clustervirial}.
We give here an improved parametrization of the DD2
 model \cite{clustervirial} in form of a Pad\'e approximation,see also \cite{R}.
The variables are temperature $T$, baryon number
density $n=n_n+n_p$, and
the asymmetry parameter  $\delta=1-2Y_p$  with the total proton fraction $Y_p=n_p/n$.
The intended relative accuracy in the parameter value range $T < 20$ MeV, $n< 0.15$ fm$^{-3}$ 
is 0.001.

The scalar  self-energy (identical for neutrons and
 protons) is approximated as
\begin{equation}
\label{scalar}
	S(T,n_B, \delta) = \frac{s_1(T,\delta) \;n + s_2(T,\delta) \;n^2
+  s_3(T,\delta)\;n^3}{1 + s_4(T,\delta)\;n + s_5(T,\delta)\; n^2}
\end{equation}
with coefficients
\begin{eqnarray}
&& s_i(T,\delta) = s_{i,0}(\delta) + s_{i,1}(\delta)\; T +  s_{i,2}(\delta)\;T^2  ,  \nonumber \\
&& s_{i,j}(\delta) = s_{i,j,0}+  s_{i,j,2}\;\delta^2 +  s_{i,j,4}\;\delta^4 ;
\end{eqnarray}
 baryon number densities $n$ in fm$^{-3}$ and temperatures $T$ as well as the self energies $S, V_\tau$ in MeV. Parameter
values are given in
Table \ref{tab:S}.

\begin{table}[ht]
\begin{tabular}{|c|c|c|c|c|c|c|}
\hline
{} & $s_{i,j,k}$& $i=1$	& $i=2$		& $i=3$		& $i=4$		& $i=5$ \\
\hline
{}& $k=0$  & 4462.35	& 204334	&  125513	&  49.0026	&   241.935	\\
$j=0$& $k=2$	&  1.63811 &  -11043.9	&  -64680.5	&  -1.76282	&  -19.8568	 \\
{} & $k=4$  & 	0.293287&  -46439.7	 &  -4940.76	&  -10.6072	&  -48.3232	\\
\hline
{}& $k=0$  & -7.22458	&  7293.23	&   1055.3	&   1.70156	&   6.6665	\\
$j=1$& $k=2$	& 0.92618 & -49220.9	 &  -19422.6	&   -11.1142	&   -52.6306	 \\
{} & $k=4$  &-0.679133 	&  35263	&  15842.8	&   7.92604	&   38.1023	\\
\hline
{}& $k=0$  &  0.00975576  &	-209.452  &   132.502	&   -0.0456724	&   -0.112997	\\
$j=2$& $k=2$	&  -0.0355021 &	 2114.07  &  572.292	&   0.473553	&   2.15092	 \\
{} & $k=4$  & 	0.026292&  -1507.55	 &   -555.762	&   -0.337016	&   -1.57597	\\
\hline
\end{tabular}
\caption{\label{tab:S}%
Coefficients $s_{i,j,k}$ for the Pad\'e approximation of the scalar
 self-energy $S(T,n,\delta)$.}
\end{table}

The vector  self-energy $V_p(T,n, \delta)=V_n(T,n, -\delta)$ is
approximated as
\begin{equation}
\label{vector}
	V_p(T,n, \delta) = \frac{v_1(T,\delta) \;n + v_2(T,\delta) \;n^2
+  v_3(T,\delta)\;n^3}{1 + v_4(T,\delta)\;n + v_5(T,\delta)\; n^2}
\end{equation}
with coefficients
\begin{eqnarray}
&& v_i(T,\delta) = v_{i,0}(\delta) + v_{i,1}(\delta)\; T +  v_{i,2}(\delta)\;T^2  ,  \nonumber \\
&& v_{i,j,k}(\delta) = v_{i,j,0}+  v_{i,j,1}\;\delta+  v_{i,j,2}\;\delta^2+  v_{i,j,3}\;\delta^3 +  v_{i,j,4}\;\delta^4 \,.
\end{eqnarray}
Parameter
values are given in
Table \ref{tab:V}.

\begin{table}[ht]
\begin{tabular}{|c|c|c|c|c|c|c|}
\hline
{} & $v_{i,j,k}$& $i=1$	& $i=2$		& $i=3$		& $i=4$		& $i=5$ \\
\hline
{}& $k=0$  & 3403.94		& -345.863 	&33553.8&2.7078		&18.7473	\\
{}& $k=1$  & -490.15		&  1521.62 	&4298.76&-0.162553	&4.0948364	\\
$j=0$& $k=2$ &  -0.0213143 	&  -2658.72	 &3692.23&-0.308454	&-0.0308012	\\
{}& $k=3$  & 0.00760759		&  -408.013	&-1083.14&-0.174442	&-0.751981	\\
{} & $k=4$  & 	0.0265109	&   -132.384	&-728.086&-0.0581052	&-0.585746	\\
\hline
{}& $k=0$  & -0.000978098	& 29.309	&-192.395&0.0161456	&-0.102959	\\
{}& $k=1$  & -0.000142646	&  -8.80748	&-52.0101&-0.00145171	&-0.044524	\\
$j=1$& $k=2$	&  0.00176929 	& -236.029	 &-141.702&-0.0689643	&-0.308021	\\
{}& $k=3$  & 0.00043752		& 13.7447	&-57.9237&-0.0000398794	&-0.0190921	\\
{} & $k=4$  & 	-0.00321724	&  111.538	&-11.4749&0.0317996	&0.0869529	\\
\hline
{}& $k=0$  & 0.0000651609	& 3.63322	&15.2158&0.00105179	&0.0118049	\\
{}& $k=1$  & 0.0000098168	& 0.0163495	&3.86652&0.000192765	&0.0021141	\\
$j=2$& $k=2$	&  -0.0000394036 & 6.88256 	 &-0.785201&0.00203728	&0.0070548	\\
{}& $k=3$  & 0.0000381407	&-0.369704	&1.59625&0.00000561467	&0.000565564	\\
{} & $k=4$  & 	0.000110931	& -3.28749	&2.0419	&-0.000932046	&-0.00182714	\\
\hline
\end{tabular}
\caption{\label{tab:V}%
Coefficients $v_{i,j,k}$ for the Pad\'e approximation of the vector
  self-energy $V_p(T,n_B,\delta)=V_n(T,n, -\delta)$.}
\end{table}

 From the density EoS (\ref{neos}) the solution  $\mu^{\rm RMF}_\tau(T,n_{\tau'})$ is found  in RMF approximation.
 From this, we can extract  the corresponding Landau parameters.
For the chemical potentials (at $T=0$ equal to the Fermi energy) we get
\begin{equation}
\mu^{\rm RMF}_\tau= [(m_\tau - S(n,T,n_{\tau'}))^2 + p_{{\rm F},\tau}^2]^{1/2} +V_\tau(n,T,n_{\tau'}).
\end{equation}

With (\ref{RMF2}), for neutron matter ($Y_p=0$) the relation
\begin{equation}
f^{\rm neutr}_0(n)= \frac{3 E^{\rm RMF}({\bf p}_{{\rm F},n};T,n,0)}{ (3 \pi^2 n)^{2/3}}n \frac{\partial \mu}{\partial n}-1
\end{equation}
is easily calculated. No instability is found.

For symmetric matter ($Y_p=0.5$) we have
\begin{equation}
f^{\rm symm }_0(n)= \frac{3 E^{\rm RMF}({\bf p}_{{\rm F},n};T,n,0.5)}{ (3 \pi^2 n/2)^{2/3}}n \frac{\partial \mu}{\partial n}-1.
\end{equation}
The chemical potentials coincide with the Fermi energy, $\mu = E^{\rm RMF}({\bf p}_{{\rm F},n};T,n,Y_p)$. Note that the Fermi momentum $p_{{\rm F},\tau} =(3 \pi^2 n_\tau)^{1/3}$ is related to the baryon density as $n_n=n$ for neutron matter, and $n_n=n/2$ for symmetric matter.

\end{document}